\documentclass[final,3p,times]{elsarticle}

\usepackage{amssymb}

\usepackage{amsmath}
\usepackage{algorithm}
\usepackage{algorithmic}
\usepackage{graphicx}
\usepackage{caption}
\usepackage{subcaption}
\usepackage{siunitx}

\usepackage{color}
\usepackage{amsmath}
\usepackage{amssymb}
\usepackage{bm}

\newcommand{\dt}{\Delta t}
\newcommand{\tvect}[1]{\MakeUppercase{\bf #1}}
\newcommand{\svect}[1]{\MakeUppercase{#1}}
\usepackage[ruled,algo2e,linesnumbered,algonl]{algorithm2e}
\usepackage{mdwlist}

\newcommand{\Qmat}{{\bm{Q}}}

\newcommand{\Uvec}{{\bm{U}}}
\newcommand{\Fvec}{{\bm{F}}}

\newcommand{\rvec}{{\bm{r}}}
\newcommand{\tauvec}{{\bm{\tau}}}

\journal{Parallel Computing}

\begin{document}

\begin{frontmatter}



\title{Toward fault-tolerant parallel-in-time integration with PFASST}


\author[add1]{Robert Speck\corref{cor1}}
\ead{r.speck@fz-juelich.de}
\author[add2]{Daniel Ruprecht}
\ead{d.ruprecht@leeds.ac.uk}
\cortext[cor1]{Corresponding author}

\address[add1]{J\"ulich Supercomputing Centre, Forschungszentrum J\"ulich GmbH, Germany}
\address[add2]{School of Mechanical Engineering, University of Leeds, Woodhouse Lane, Leeds LS2 9JT, UK}

\begin{abstract}
We introduce and analyze different strategies for the parallel-in-time integration method PFASST to recover from hard faults and subsequent data loss.
Since PFASST stores solutions at multiple time steps on different processors, information from adjacent steps can be used to recover after a processor has failed.
PFASST's multi-level hierarchy allows to use the coarse level for correcting the reconstructed solution, which can help to minimize overhead.
A theoretical model is devised linking overhead to the number of additional PFASST iterations required for convergence after a fault.
The potential efficiency of different strategies is assessed in terms of required additional iterations for examples of diffusive and advective type.
\end{abstract}


\end{frontmatter}

The extremely high number of cores in today's and future supercomputing architectures leads to a wide range of challenges that developers of numerical methods have to face.
Most obvious is probably the requirement for algorithms to offer a maximum degree of concurrency: numerical methods with strong serial dependencies will never perform well on massively parallel computers.
Probably equally important are the complications arising from decreasing \emph{mean-times between failures} (MTBF).
As computers feature more and more hardware components, the probability of one component failing during a simulation increases.
Leadership supercomputers already experience a MTBF of a couple of hours~\cite{DongarraEtAl2013} and the massive increase in components on the path to exascale computing will greatly exacerbate this problem~\cite{IESR}.
Moreover, relaxing reliability on the hardware side and shifting more and more responsibility to deal with faults to the application is a possible way to reduce energy consumption -- provided that the used software can properly deal with faults.
Therefore, well-developed strategies to deal with faults on the side of numerical methods can help to reduce energy-to-solution.
The central significance of fault tolerance for extreme scale computing has been widely recognized.
A recent overview is given for example by Cappello et al.~\cite{CappelloEtAl2014}.

We adopt the nomenclature proposed by Snir et al.~\cite{SnirEtAl2014}: faults occur at the system level and can cause errors.
Errors may then lead to failures when causing transition to an incorrect system state. 
Some faults can cause the system to crash but others will just corrupt the state and cause the system to return a wrong solution (''silent errors'').
Faults that are transient in nature are referred to as soft or elusive errors while reproducible errors are called hard or solid.
\emph{Fault tolerance} refers to the capacity to detect faults and to apply contingency procedures to bring the system back into a correct state.

The most straightforward approach is checkpointing combined with a backward recovery strategy: here, the state of the system is frequently saved and, should a failure occur, the simulation is rolled back to the last correct state and restarted from there.
Simple restarting causes massive overhead and is likely not a feasible strategy for exascale systems, at least not on its own~\cite{BosilcaEtAl2013}.
Much attention has therefore been paid to \emph{algorithm-based fault tolerance} (ABFT) strategies which exploit specific features of employed numerical methods for forward recovery.
Upon detection of a fault, the application proceeds with recovery steps to correct or retrieve (sometimes partially) lost data and bring the system back into a correct state.
The concept was first studied for soft errors in matrix operations~\cite{HuangEtAl1984} but has since then been investigated for a wide range of iterative algorithms, for example in the field of numerical linear algebra, and also been applied to hard faults~\cite{ChenDongarra2008,Chen2011,YaoEtAl2012}.

In this paper we consider resilience against compute node hard errors and failures, typically related to the failure of some hardware component, for applications solving time-dependent partial differential equations.
With a distributed memory paradigm, typically MPI, a failed node results in the loss of data stored on that node.
Usually, this causes a job to either crash or stall.
While some MPI libraries support coordinated checkpointing without involvement of the application, routines that allow for \emph{user-level failure mitigation} (ULFM) and specifically algorithm-based fault tolerance are not yet part of the MPI standard~\cite{BlandEtAl2014}.
Implementation of ABFT techniques into an MPI code would therefore require using experimental extensions of MPI, which are subject of current research.

Parallel-in-time integration methods have been mainly considered as means to extend strong scaling limits of spatial parallelization and/or to improve utilization of very large machines~\cite{SpeckEtAl2012,RuprechtEtAl2013_SC}.
However, ''parallel across-the-steps'' methods like Parareal~\cite{LionsEtAl2001}, PITA~\cite{FarhatEtAl2003} (''parallel implicit time-integrator'') or PFASST~\cite{EmmettMinion2012} (''parallel full approximation scheme in space and time'') share features that make them natural candidates for algorithmic-based fault tolerance: (i) they hold copies of the (approximate) solution at different times on different processes, (ii) they are iterative, and (iii) they use a level hierarchy with at least one computationally cheap, coarse level.
Therefore, should a process fail and the solution at one point in time is lost, a recovery process can retrieve approximate values from the previous and following processes in time.
This reconstructed solution can then be improved by iterating on the cheap coarse level, causing minimal overhead.
After recovery, continuing with the iteration leads to a solution of the same accuracy at the end -- although probably at the cost of additional iterations and thus more work.
Because time stepping is typically the outermost loop for the numerical solution of a time-dependent PDE, protecting it by ABFT covers a larger area of the code than if only e.g. the linear or nonlinear spatial solver is protected.
However, to the best of our knowledge, no publications exist so far that consider ABFT for parallel-in-time integration.\footnote{Shortly after submission of this article, a preprint has been published independently that provides a similar analysis for the Parareal algorithm~\cite{NielsenHesthaven2016}.}

In this paper, we propose a recovery strategy from hard faults for the PFASST method and demonstrate its efficacy.
For spectral deferred corrections, the serial time integrator at the heart of PFASST, resilience against soft faults has been demonstrated before~\cite{MinionEtAl2015_ftsdc}.
Even though the iterative nature of PFASST suggests that it should also provide some resilience against soft faults, in this paper we consider only hard failures, where a process crashes and information handled by this process is lost completely.
Since ULFM is not yet part of the MPI standard, we do not report runtimes from an actual MPI implementation but assess overhead in terms of additional PFASST iterations which, through a simple theoretical model, can be directly linked to computational cost.
The software used for the numerical examples presented in this paper is publicly available~\cite{pySDC2015} to allow for the reproduction and extension of presented results.

A forward recovery strategy for a spatial multi-grid solver has recently been proposed~\cite{RuedeEtAl2015}.
There, upon failure-stop of a process holding a sub-domain in a distributed memory multi-grid solver, information from processes holding adjacent domains is used to reconstruct the lost information.
Based on the parallel-in-time solver PFASST, we propose a similar strategy for the time axis.
Ultimately, interweaving parallel-in-time integration with iterative spatial solvers may not only be worthwhile to optimize efficiency~\cite{Mula2014,MinionEtAl2015} but also provide a promising direction for the construction of fault-tolerant methods for the integration of time-dependent partial differential equations.

\section{The PFASST algorithm}
The PFASST algorithm has been introduced by Emmett and Minion in 2012~\cite{EmmettMinion2012}.
It is based on an extension of spectral deferred corrections~\cite{DuttEtAl2000} (SDC) to a multi-level hierarchy plus the incorporation of a \emph{full approximation scheme} correction term on the coarse levels.
These \emph{multi-level SDC} iterations~\cite{SpeckEtAl2015_BIT} are run concurrently on multiple time steps with each time step frequently sending forward updated initial values.
In the following, we briefly describe SDC, MLSDC and PFASST, focusing on the key aspects which are of interest for the work presented here.
Detailed descriptions of SDC, MLSDC and PFASST are available elsewhere~\cite{EmmettMinion2012,SpeckEtAl2015_BIT,Minion2003}.

\subsection{Spectral deferred corrections (SDC)}
Consider an initial value problem on a single time step $[T_n, T_{n+1}]$ in integral form
\begin{equation}
	\label{eq:picard}
	u(T_{n+1}) = u(T_n) + \int_{T_n}^{T_{n+1}} f(u(t),t)~dt
\end{equation}
for $u(t), f(u(t),t)\in\mathbb{R}^N$, where $N\in\mathbb{N}$ are e.g.~the number of degrees-of-freedom in spatial dimensions.
For simplicity, we give the derivation for $N=1$ and $f(u,t) = f(u)$, but the extension to systems of initial value problems stemming e.g.~from a method of lines approach for PDEs as well as to non-autonomous problems is straightforward.
Let $T_{n} \leq t_1 < \ldots < t_M \leq T_{n+1}$ denote a set of $M$ quadrature nodes within $[T_n, T_{n+1}]$.
Using a quadrature rule like Gauss-Radau or Gauss-Lobatto to approximate the integral in~\eqref{eq:picard} provides the equations 
\begin{equation}
	U_m = U_0 + \Delta t \sum_{j=1}^{M} q_{m,j} f(U_j, t_j), \ \text{for} \ m=1,\ldots,M,
\end{equation}
for the approximate solutions $U_m \approx u(t_m)$ with $U_0 = u(T_n)$.
The $q_{m,j}$ are quadrature weights obtained by integrating Lagrange polynomials.
The $U_m$ are equivalent to the stages of an implicit Runge-Kutta methods~\cite[Theorem 7.7]{hairer_nonstiff} and such methods are referred to as \emph{collocation methods}.
To compute these stages, a large, fully implicit system of equations of the form
\begin{equation}\label{eq:collocation}
  \Uvec = \Uvec_0 + \Delta t\, \Qmat\, \Fvec(\Uvec)
\end{equation}
has to be solved, where $\Uvec_0 := \bm{I}_M\otimes U_0$ is the vector of initial values, $\Qmat := (q_{m,j})_{m,j=1,...,M}$ is the so-called quadrature matrix and $\Fvec(\Uvec) := \left(f(U_1),...,f(U_M)\right)^T $ is the vector of function values.
Instead of solving this system directly using e.g. a Newton-Raphson method, SDC provides an efficient preconditioned iterative approach which converges to the solution $\Uvec$.
This iteration can be formulated on a node-to-node basis, where each iteration updates the solutions at the nodes one by one.
For $\Qmat_j$ being the $j$th row of $\Qmat$ we define
\begin{equation}
    \Delta t S_m \Fvec(\Uvec^k) := \Delta t (\Qmat_{m+1} - \Qmat_m)\Fvec(\Uvec^k) \approx \int_{t_m}^{t_{m+1}} f(U^k(s),s)\mathrm{d} s
\end{equation}
as an approximation to the node-to-node integral of $f$ at iteration $k$.
Omitting the detailed derivation, an SDC sweep in node-to-node formulation with implicit Euler as preconditioner reads
\begin{equation}\label{eq:sdc_sweep}
	U^{k+1}_{m+1} = U^{k+1}_{m} + \Delta t_m \left(f(U_{m+1}^{k+1}) - f(U_{m+1}^k)\right) + \Delta t S_m\Fvec(\Uvec^k),\quad m = 0,...,M-1.
\end{equation}
The preconditioner can easily be replaced by other time-stepping methods like explicit Euler or even higher order methods and IMEX schemes. 
In the latter case, the resulting \emph{semi-implicit SDC}~\cite{Minion2003} (SISDC) can treat stiff parts of the right-hand side $f$ implicitly and non-stiff parts explicitly. 
In each iteration~\eqref{eq:sdc_sweep}, $M$ (possibly non-linear) equations of the form $U^{k+1}_{m+1} - \dt f(U_{m+1}^{k+1}) = b_m^k$ need to be solved while the original system~\eqref{eq:collocation} requires the solution of a single (possibly non-linear) system of size $M \times M$ .
The iteration is stopped when the residual
\begin{equation}
    \rvec^k = \Uvec_0 + \Delta t \Qmat \Fvec(\Uvec^k) - \Uvec^k
\end{equation}
falls below a predefined tolerance or a set maximum number of iterations $K_\text{max}$ is reached.

\subsection{Multi-level SDC (MLSDC)}

The goal of MLSDC~\cite{SpeckEtAl2015_BIT} is to replace some of the costly fine level sweeps with sweeps on coarser and cheaper levels of a space-time hierarchy.
To this end, system~\eqref{eq:collocation} is solved using ideas from nonlinear multigrid methods. 
With SDC sweeps playing the role of a ''smoother'' on each level of the hierarchy, a V-cycle is performed.
The systems on the coarse levels are augmented with a $\tau$-correction which corresponds to the correction term of the multigrid full approximation scheme (FAS)~\cite{brandt:1977}.
Using two levels for simplicity, this $\tau$-correction reads
\begin{equation}
\tauvec^k = \dt\left(R\Qmat_\mathrm{f}\Fvec_\mathrm{f}(\Uvec^k_\mathrm{f}) - \Qmat_\mathrm{c}\Fvec_\mathrm{c}(R\Uvec^k_\mathrm{f})\right),
\end{equation}
where $R$ is a space-time restriction operator and the indices f and c refer to the fine and coarse versions of the quadrature matrix, the function values and the solution values.
The original SDC sweep~\eqref{eq:sdc_sweep} is modified on the coarse level by adding a $\tau$-correction term, so that
\begin{equation}\label{eq:sdc_sweep_c}
	U^{k+1}_{\mathrm{c},m+1} = U^{k+1}_{\mathrm{c},m} + \Delta t_m \left(f_\mathrm{c}(U_{\mathrm{c},m+1}^{k+1}) - f_\mathrm{c}(U_{\mathrm{c},m+1}^k)\right) + \Delta t S_{\mathrm{c},m}\Fvec(\Uvec^k) + \tauvec^k_{m+1}-\tauvec^k_{m}.
\end{equation}
These values are then used to apply a coarse level correction to the values on the fine level with
\begin{equation}
    \Uvec_{\mathrm{f}}^{k+1} = \Uvec_{\mathrm{f}}^{k+1} + P(\Uvec_{\mathrm{c}}^{k+1} - R\Uvec_{\mathrm{f}}^{k}),
\end{equation}
where $P$ is a space-time prolongation operator and indices f and c indicate fine and coarse level again.
The key to an efficient MLSDC scheme is choosing suitable coarsening strategies.
With FAS helping to represent information of the fine level on coarser ones, the classical strategies are~\cite{SpeckEtAl2015_BIT}:
\begin{itemize*}
    \item reduction of quadrature nodes (reduced order of time discretisation),
    \item reduction of degrees-of-freedom in space (especially for PDE-based ODE systems),
    \item reduced order of the space discretisation,
    \item inexact implicit solves if implicit SDC is used,
    \item reduced physical model of the problem.
\end{itemize*}

\subsection{Parallel full approximation scheme in space and time (PFASST)}

In order to parallelize in time, PFASST initiates MLSDC cycles on multiple time steps on different processors, with frequent communication of updated initial values on all levels.
The key is that closely synchronized, blocking communication is required only at the coarsest and cheapest level.
All other levels have weaker dependencies and allow for substantial overlapping of computation and communication, in particular when using more than two levels.

\begin{figure}[!ht]
	\centering
    \includegraphics[scale=1]{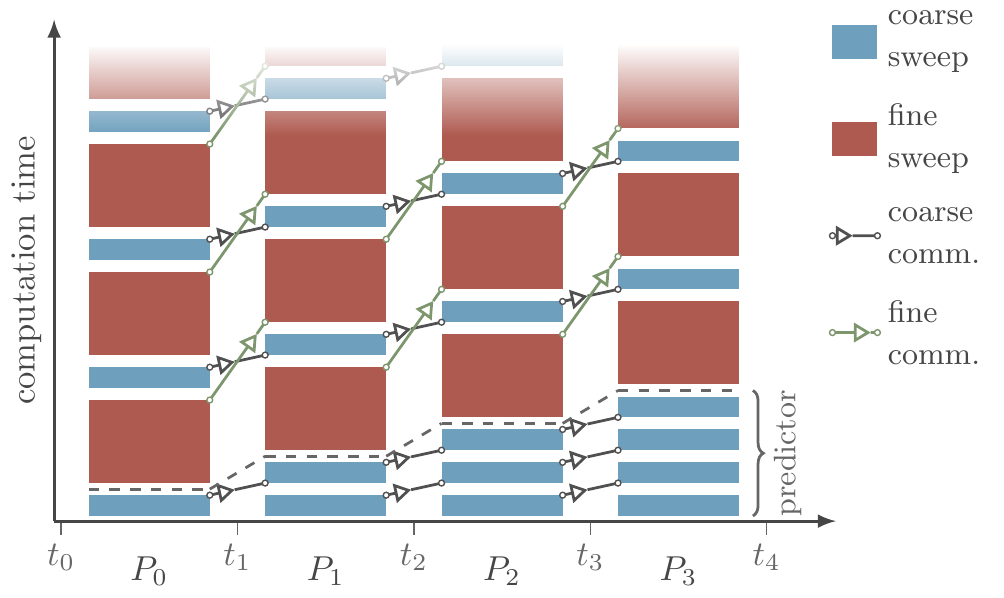}
	\caption{Schematic view of the PFASST algorithm with two levels and four processes $P_0,...,P_3$ handling four parallel time steps. Only communication on the coarsest level is blocking, communication on other levels can be overlapped with computation. Created using pfasst-tikz~\cite{pfasst-tikz}.}
	\label{fig:pfasst}
\end{figure}

A simple example is sketched in Figure~\ref{fig:pfasst}: four parallel time steps are shown, each with two levels.
It is often beneficial to perform a start-up phase (``predictor''), where the initial value $U_0$ is communicated to all processes and a number of coarse level sweeps is performed to produce a very early guess of the solution~\cite{EmmettMinion2012}.
After the prediction phase, the processes $P_0, ..., P_3$ perform their fine sweep (large red blocks) simultaneously and send the result at $t_1, ..., t_4$ forward in time. 
This communication is non-blocking and all processes can directly continue with the iteration.
This leads to a staggered or pipelined execution if communication and computation are correctly overlaid~\cite{EmmettMinion2014_DDM}.
In the next step, the solution from the fine level is restricted, the $\tau$-correction is formed and the coarse sweep (blue small blocks) is started.
For the coarse sweep, each process has to wait for the new initial condition to arrive from the previous process, so sweeps on the coarsest level are performed in serialized order using blocking communication.
After receiving an updated fine level initial value from the previous process, the coarse level correction is computed and added to the fine solution.
This approach can easily be extended to multiple levels and an algorithmic view of PFASST with $L+1$ levels is provided in Algorithm~\ref{alg:pfasst}.
As for SDC and MLSDC, the iteration stops if the residual on all time steps falls below a predefined threshold or if the maximum number of PFASST iterations $K_\text{max}$ is reached.
Typically, the last process in time will have to perform the most iterations and determine the overall runtime.

\begin{algorithm2e}[!th]
	\caption{PFASST iteration}
	\label{alg:pfasst}
	 \SetKwComment{Comment}{\# }{}
	 \SetKwInOut{Input}{input}
	 \SetKwInOut{Output}{output}  	 
	\Input{Values $U^{k}_{m,l}$ for $m=1,\ldots,M$ and $l=0,\ldots,L$.}
	\Output{Update values $U^{k+1}_{m,l}$.}

	\Comment{On fine level}
	Perform SDC sweep on finest level $0$ with $\Uvec_0$, $\Fvec_0$\\
	Send updated final value $U_{M,l}$ to following process (non-blocking)\\
	
	\Comment{Move from fine ($l=0$) to coarsest level ($l=L$)}
	Restrict values from level $0$ to level $1$\\
	Compute FAS correction $\tau_1$\\
	\For{$l=1,L-1$}{
		
		Perform SDC sweep with $\Uvec_l$, $\Fvec_l$, $\tau_l$\\
		Send updated final value $U_{M,l}$ to following process (non-blocking)\\
		Restrict values from level $l$ to level $l+1$\\
		Compute FAS correction $\tau_{l+1}$\\
	}
	
	\Comment{On coarsest level}
	Receive new initial value $U_{0,L}$ on coarsest level from previous process\\
	Perform SDC sweep with $\Uvec_L$, $\Fvec_L$, $\tau_L$\\
	Send updated final value $U_{M,L}$ to following process (blocking)\\
	
	\Comment{Move from coarsest to fine level}
	\For{l=L-1,1}{
    	Receive new initial value $U_{0,l}$ for current level from previous process\\
    	Interpolate and apply coarse correction from level $l+1$ to $\Uvec_l$\\	
		Perform SDC sweep with $\Uvec_l$, $\Fvec_l$, $\tau_l$\\
	}
	
	\Comment{On fine level}
	Receive new initial value $U_{0,0}$ for current level from previous process\\
	Interpolate and apply coarse correction from level $1$ to $\Uvec_0$\\	

\end{algorithm2e}

\section{Fault-tolerant PFASST}
This section introduces different strategies to recover solutions lost due to failure of a process.
Since in PFASST each process handles one time step, we assume here that a fault leads to the loss of all data associated with a specific time step.
Without a recovery strategy, the whole block of parallel time steps will have to be computed again after a fault occurs, discarding the information obtained so far (backward recovery).
Clearly, when using an iterative method like PFASST, this is the least efficient way of treating failures.
The presented recovery strategies use information from adjacent time steps to reconstruct the lost data on a replacement process (forward recovery).
For a discussion of this replacement process, we refer to Section~\ref{sect:outro}.
For now, we assume that we can either use the failed process again (e.g.~after a reboot) or that there is a spare replacement process available.
We furthermore assume that a fault occurs right before a fine sweep or, equivalently, at the end of an iteration. 
Therefore, all recovery strategies have to provide enough information to continue with a fine sweep at the beginning of a V-cycle.
Of course, in a real-world simulation, recovery strategies will have to deal with faults occurring at any phase in the algorithm.
Restarting a failed step at the beginning of a V-cycle is therefore a reasonable choice, because enough information is available for the adjacent processes to continue their cycles with information on all levels.
Also, as soon as the replacement process reaches the coarse level during the V-cycle, the blocking communication ``aligns'' this process and its computation with the workflow of the others.

\subsection{Recovery strategies}\label{sect:recovery_strategies}

Exploiting the iterative and multi-level nature of PFASST, we present four different strategies which allow a re-use of data from pre- and succeeding  time steps for the recovery of lost data: one-sided interpolation with and without coarse correction and two-sided interpolation with and without coarse correction.

\subsubsection{One-sided interpolation}
The one-sided recovery strategy reconstructs lost information using data from the process handling the previous time step.
In its most simple form, the replacement process fetches the current initial value on the finest level from the previous time step (which is the final value there) and spreads it to all quadrature nodes, corresponding to constant interpolation.
Using these reconstructed fine level values, the standard MLSDC V-cycle is restarted to populate the coarse level and continue the PFASST iteration.
Optionally, before starting the V-cycle, an additional coarse correction step can be performed, see Section~\ref{subsubsec:coarse-correct}.
The coarse level is populated using the restriction operator $R$ and values are improved by performing a set number $n_\text{rec}$ of coarse level recovery sweeps.
Coarse level values are then used to correct the fine level values before restarting the MLSDC V-cycle.
Numerical experiments in Section~\ref{sec:experiments} show that the coarse correction step can greatly reduce the overhead in terms of required additional iterations.
This recovery strategy is sketched in Algorithm~\ref{alg:one-sided-recovery}.
\begin{algorithm2e}[!h]
	\caption{1-sided recovery after failure of process $p$ in iteration $k$.}\label{alg:one-sided-recovery}
  	\SetKwComment{Comment}{\# }{}
	\SetCommentSty{textit}
	\SetKwInOut{Input}{input}
         \SetKwInOut{Output}{output}
        \Input{rank $p$, flag \texttt{do\_correction}, number of recovery sweeps $n_{\text{rec}}$}
        \Output{reconstructed fine and coarse level values $\tvect{U}^{k,\text{f}}$, $\tvect{U}^{k,\text{c}}$}	
	\Comment{Receive new starting value from previous process}
	Receive $\svect{u}_0$ from process $p-1$\\
	\Comment{Spread new initial value to all fine level nodes and recompute $f$}
	\For{$m=0,M_{\text{f}}-1$}{
		$\svect{u}^{k,\text{f}}_m \leftarrow \svect{u}_0$\\
		Compute $f(\svect{u}^{k,\text{f}}_m)$
	}
   	\Comment{Perform possible additional coarse-level corrections}
	\uIf{\textup{\texttt{do\_correction}}}{ $\tvect{U}^{k,\text{f}}, \tvect{U}^{k,\text{c}} \leftarrow$ coarse-level-correction($\tvect{U}^{k,\text{f}}, n_{\text{rec}}$) }
	\uElse{$\tvect{U}^{k,\text{c}} \leftarrow 0$}
	Start MLSDC iteration $k+1$ on fine level
\end{algorithm2e}

\subsubsection{Two-sided interpolation}
Instead of using only information from the previous time step, additional information from the following time step can be used to reconstruct lost information.
This strategy uses linear interpolation to repopulate the fine level nodes.
As for the one-sided strategy, using coarse level corrections before restarting the MLSDC V-cycle can greatly reduce the number of additional iterations required.
This strategy is sketched in Algorithm~\ref{alg:two-sided-recovery}.
In the examples analyzed in Section~\ref{sec:experiments}, two-sided recovery produces slightly better results than one-sided recovery.
However, it requires one additional message to be send during recovery.
\begin{algorithm2e}[!h]
	\caption{2-sided recovery after failure of process $p$ in iteration $k$.}\label{alg:two-sided-recovery}
  	\SetKwComment{Comment}{\# }{}
	\SetCommentSty{textit}	
	\SetKwInOut{Input}{input}
         \SetKwInOut{Output}{output}
        \Input{rank $p$, flag \texttt{do\_correction}, number of recovery sweeps $n_{\text{rec}}$}
        \Output{reconstructed fine and coarse level values $\tvect{U}^{k,\text{f}}$, $\tvect{U}^{k,\text{c}}$}
	\Comment{Receive starting value from previous process}
	Receive $\svect{u}_0$ from process $p-1$\\
	\Comment{Receive end value from following process}
	Receive $\svect{u}_{\text{end}}$ from process $p+1$\\
	\Comment{Repopulate fine level values through linear interpolation}
	\For{$m=0,M_{\text{f}}-1$}{
		$\svect{u}^{k,\text{f}}_m \leftarrow \left(1  - \tau_m \right) \svect{u}_{\text{end}} + \tau_m \svect{u}_{0}$\\
		Compute $f(\svect{u}^{k,\text{f}}_m)$		
	}
    	\Comment{Perform possible additional coarse-level corrections}
	\uIf{\textup{\texttt{do\_correction}}}{ $\tvect{U}^{k,\text{f}}, \tvect{U}^{k,\text{c}} \leftarrow$ coarse-level-correction($\tvect{U}^{k,\text{f}}, n_{\text{rec}}$) }
	\uElse{$\tvect{U}^{k,\text{c}} \leftarrow 0$}
	Start MLSDC iteration $k+1$ on fine level	
\end{algorithm2e}

\subsubsection{Coarse level corrections}\label{subsubsec:coarse-correct}
One-sided as well as two-sided recovery can be augmented by performing additional coarse level sweeps before restarting the local V-cycles.
This provides a coarse level correction to the reconstructed fine level values, increases the accuracy of the reconstructed solution and helps to reduce the number of additional iterations required by PFASST to converge after a fault.
Coarse level corrections entail the cost of a number of sweeps on the coarse level.
For all experiments documented here, we sweep on the coarse level until either the number of sweeps $n_\text{rec}$ is as large as the number of iterations before the failure or the residual of the coarse level is below the residual of the coarse level on the previous time step.
Both numbers need to be received from the previous process as part of the recovery process.
Also, the maximal number of sweeps is limited by the number of parallel time-steps $P$, see Section~\ref{subsec:overhead}.
Since PFASST can only be efficient when coarse level sweeps are cheap compared to sweeps on the fine level, the resulting overhead should be relatively small, see also Section~\ref{subsec:overhead}.
The proposed coarse level correction strategy is sketched in Algorithm~\ref{alg:coarse-correction}.
For simplicity, we sketch the coarse level correction only for two-level PFASST.
For multi-level PFASST, this correction would be performed on the coarsest level only in order to minimize overhead.
This approach is similar to the standard predictor phase in the PFASST algorithm.
\begin{algorithm2e}[!h]
	\caption{coarse level correction}\label{alg:coarse-correction}
  	\SetKwComment{Comment}{\# }{}
	 \SetKwInOut{Input}{input}
         \SetKwInOut{Output}{output}
	\Input{fine level values $\tvect{U}^{k,\text{f}}$, number of recovery sweeps $n_{\text{rec}}$}
	\Output{updated coarse values $\tilde{\tvect{U}}^{k,\text{c}}$, updated fine level values $\tilde{\tvect{U}}^{k,\text{f}}$}
	\SetCommentSty{textit}		
	\Comment{Repopulate coarse level through restriction}
	Restrict $\tvect{U}^{k,\text{c}} \leftarrow \tvect{U}^{k,\text{f}}$\\
	\Comment{Perform additional coarse sweeps}
	$\tilde{\tvect{U}}^{k,\text{c}} \leftarrow \tvect{U}^{k,\text{c}}$\\
	\For{$k=1,n_{\text{rec}}$}{
		coarse sweep update of $\tilde{\tvect{U}}^{k,\text{c}}$
	}
	\Comment{Compute updated coarse correction}
	$\tilde{\tvect{U}}^{k,\text{f}} \leftarrow \tvect{U}^{k,\text{f}} + \text{Interpolate}\left( \tilde{\tvect{U}}^{k,\text{c}} - \tvect{U}^{k,\text{c}}\right)$\\
	\Return $\tilde{\tvect{U}}^{k,\text{f}}$, $\tilde{\tvect{U}}^{k,\text{c}}$
\end{algorithm2e}

\subsection{Overhead and efficiency}\label{subsec:overhead}
Overhead is a key metric to assess the performance of strategies for resilience.
It is defined as the difference between the wall clock time $T_{\text{fault}}$ of a simulation in a faulty system with recovery minus the wall clock time $T_{\text{no-fault}}$ of a simulation in an ideal fault-free system~\cite{SnirEtAl2014}
\begin{equation}
	O = T_{\text{fault}} - T_{\text{no-fault}}.
\end{equation}
We have theoretical models for both and can thus derive a theoretical model for $O$, which allows to assess the efficiency of our recovery by using the number of iterations of PFASST as a proxy.
Two-level PFASST without faults runs approximately in wall clock time
\begin{equation}
	\label{eq:wallclock_nofault}
	T_{\text{no-fault}} = P n_{\text{c}} \Gamma_{\text{c}} + K \left( n_{\text{c}} \Gamma_{\text{c}} + n_{\text{f}} \Gamma_{\text{f}} \right) = \left( P + K \right) n_{\text{c}} \Gamma_{\text{c}} + K n_{\text{f}} \Gamma_{\text{f}}.
\end{equation}
Here, $P$ is the number of processors, $K$ the number of iterations, $n_{\text{c}}$ and $n_{\text{f}}$ are the number of coarse and fine sweeps per iteration while $\Gamma_{\text{c}}$, $\Gamma_{\text{f}}$ are the cost of a coarse or fine sweep.
The model ignores communication cost and other overhead but gives a reasonable estimate of PFASST's performance~\cite{SpeckEtAl2012,RuprechtEtAl2013_SC}.
Now, if PFASST simply restarts after failure in iteration $K_{\text{fault}}$, the wall clock time is
\begin{equation}
	\label{eq:wallclock_restart}
	T_{\text{fault}}^{\text{restart}} = 2 P n_{\text{c}} \Gamma_{\text{c}} + \left( K + K_{\text{fault}} \right) \left( n_{\text{c}} \Gamma_{\text{c}} + n_{\text{f}} \Gamma_{\text{f}} \right) = \left( 2 P + K + K_{\text{fault}} \right) n_{\text{c}} \Gamma_{\text{c}} + \left( K + K_{\text{fault}} \right) n_{\text{f}} \Gamma_{\text{f}}
\end{equation}
with $K_{\text{fault}}$ being the iteration in which the fault occurs.
When PFASST performs a full restart, the number of iterations required for convergence after the fault remains the same and the total number of iterations is $K + K_{\text{fault}}$.
By taking the difference of~\eqref{eq:wallclock_nofault} and~\eqref{eq:wallclock_restart}, we can compute the overhead of a simple restarting strategy as
\begin{equation}
	O_{\text{restart}}= \left( P + K_{\text{fault}} \right) n_{\text{c}} \Gamma_{\text{c}} + K_{\text{fault}} n_{\text{f}} \Gamma_{\text{f}}.
\end{equation}
Because a fine sweep is much more expensive than a coarse sweep, that is $\Gamma_{\text{f}} \gg \Gamma_{\text{c}}$, the additional fine sweeps required by restarting will lead to significant overhead, particularly if the fault occurs late in the iteration and $K_{\text{fault}}$ is large.
For a recovery procedure that performs $n_{\text{rec}}$ coarse level sweeps, we get the following wall clock time estimate:
\begin{subequations}
\begin{align}
	T_{\text{fault}}^{\text{recovery}} 
	   &	= P n_{\text{c}} \Gamma_{\text{c}} + (K+K_{\textrm{add}}) \left( n_{\text{c}} \Gamma_{\text{c}} + n_{\text{f}} \Gamma_{\text{f}} \right) + n_{\text{rec}} \Gamma_{\text{c}} + \Gamma_{\text{rec}}  \\
	   &= \left(  P + K + K_{\textrm{add}} \right) n_{\text{c}}  \Gamma_{\text{c}} + \left( K + K_{\textrm{add}} \right) n_{\text{f}} \Gamma_{\text{f}} + n_{\text{rec}} \Gamma_{\text{c}} + \Gamma_{\text{rec}},
\end{align}
\end{subequations}
where $\Gamma_{\text{rec}}$ measures overhead from the reconstruction step due to communication of data or idle times.
Because the reconstructed values are not perfect, PFASST will typically require more iterations to converge than in the non-fault case: the number of additional iterations requires is denoted as $K_{\text{add}}$.
The overhead from recovery then is
\begin{equation}
	O_{\text{recovery}} = \left( K_{\text{add}} n_{\text{c}} + n_{\text{rec}} \right) \Gamma_{\text{c}} + K_{\text{add}} n_{\text{f}} \Gamma_{\text{f}} + \Gamma_{\text{rec}}.
\end{equation}
We define $\alpha := n_{\text{c}} \Gamma_{\text{c}}/n_{\text{f}} \Gamma_{\text{f}}$ as the coarse-to-fine ratio.
The ratio between the overhead of a full restart and the overhead of a recovery strategy then is given by
\begin{equation}\label{eq:ratio_overhead}
    \frac{O_{\text{restart}}}{O_{\text{recovery}}} = \frac{(1+\alpha) K_\text{fault} + \alpha P}{(1+\alpha)K_\text{add} + \alpha n_\text{rec}/n_\text{c} + \frac{\Gamma_{\text{rec}}}{n_{\text{f}} \Gamma_{\text{f}}}}.
\end{equation}
Clearly, an effective recovery strategy has to satisfy $O_{\text{recovery}} \leq O_{\text{restart}}$.
This requires that (i) $K_{\text{add}} \leq K_{\text{fault}}$, (ii) $n_\text{rec} \leq n_\text{c}P$ and (iii) $\Gamma_{\text{rec}} \ll n_{\text{f}} \Gamma_{\text{f}}$.
As mentioned in Section~\ref{sect:recovery_strategies}, we choose $n_\text{rec} \leq P$, so that the second criterion is always satisfied.
Since the parameter $\alpha$ controls the potential speedup achieved by the temporal parallelization with PFASST~\cite{SpeckEtAl2012,EmmettMinion2012}, it can assumed to be small: otherwise it would not make much sense to use PFASST in the first place.
The main contributions to the reconstruction overhead $\Gamma_{\text{rec}}$ will be the time required to start up a replacement process and for it to receive data.
Start-up times for a replacement process will depend on the architecture and MPI implementation and efficient strategies like ``hot replacement'' are available to keep them small~\cite{YaoEtAl2012}.
Communication costs during recovery of a single step will be equal or smaller to the communication cost of a full PFASST iteration (which requires communication for \emph{all} processes) and thus, in a regime where PFASST is reasonably effective, will be smaller than the cost of a fine sweep so that $\Gamma_{\text{rec}} \ll \Gamma_{\text{f}}$.
Therefore, the key criterion to decide whether a recovery strategy is efficient compared to a simple restart is $K_{\text{add}} < K_{\text{fault}}$.

\section{Experiments}\label{sec:experiments}
We start with parameter studies for two prototype problems in Subsection~\ref{subsec:simple_problems}.
While the problems are idealized, their simplicity allows to study a wide range of parameters in a reasonable amount of time.
Their dynamics (one diffusive, the other advective) are also representative for a wider range of more complex problems.
Performance of the proposed recovery strategies is then illustrated in Subsections~\ref{subsec:gray-scott} and~\ref{subsec:boussinesq} for two complex problems, the Gray-Scott diffusion-reaction model and the linearized Boussinesq equations for stratified, compressible flow.
All experiments are performed using the Python framework pySDC~\cite{pySDC2015}.

\subsection{Prototype problems}\label{subsec:simple_problems}
As a diffusive example problem, the one-dimensional heat equation with a forcing term is used:
\begin{equation}
    \label{eq:hard_prob_A}
    \begin{aligned}
        u_t(x,t) &= \nu u_{xx}(x,t) + f(x,t),\quad x\in[0,1],\ t\in [0,8],\\
        f(x,t) &= -\sin(\pi x)(\sin(t) - \nu\pi^2\cos(t))\quad \text{for all } x\in[0,1],\ t\in [0,8]\\
        u(0,t) &= u(1,t) = 0\quad \text{for all } t\in[0,8],\\
        u(x,0) &= \sin(\pi x)\quad \text{for all } x\in[0,1],\\
        \nu &= 0.5,\ \dt = 0.5,\ N = 255.
    \end{aligned}
\end{equation}
As an advective example, the one-dimensional transport equation is studied:
\begin{equation}\label{eq:hard_prob_B}
    \begin{aligned}
        u_t(x,t) &= c u_x(x,t),\quad x\in[0,1],\ t\in [0,2],\\
        u(0,t) &= u(1,t)\quad \text{for all } t\in[0,2],\\
        u(x,0) &= \cos(2\pi x)\quad \text{for all } x\in[0,1],\\
        c &= 1.0,\ \dt = 0.125,\ N = 256.
    \end{aligned}
\end{equation}
Here, $N$ is the number of degrees-of-freedom in space for the finite difference discretizations. 
In both cases, we use $P=16$ parallel time steps and $M=5$ Gauss-Lobatto collocation nodes for the temporal discretization.
Only one sweep is performed per level and iteration, so that $n_\text{c} = n_\text{f} = 1$.

\begin{figure}[!th]
	\centering
	\begin{minipage}[t]{0.99\textwidth}
		\centering
		\subcaption{Heat equation}
		\includegraphics[scale=1]{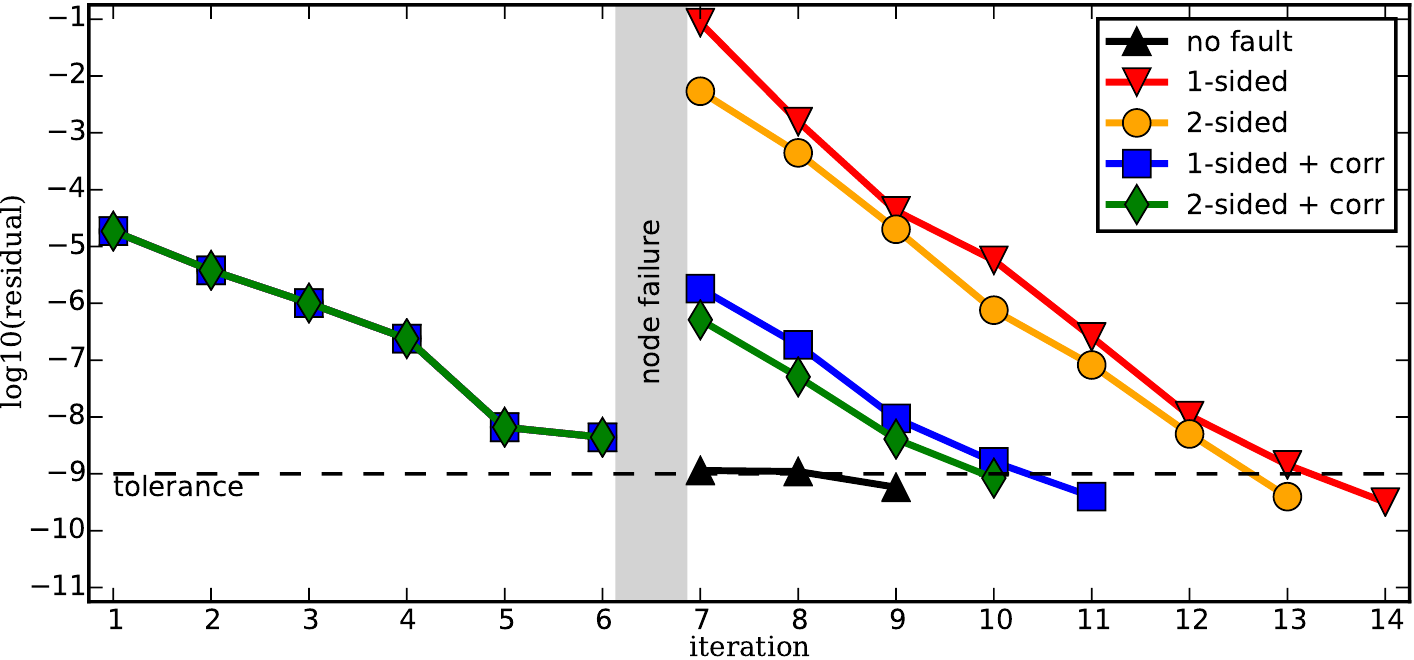}\vspace*{0.5em}
	\end{minipage}
	\begin{minipage}[t]{0.99\textwidth}
		\centering
		\subcaption{Advection equation}
		\includegraphics[scale=1]{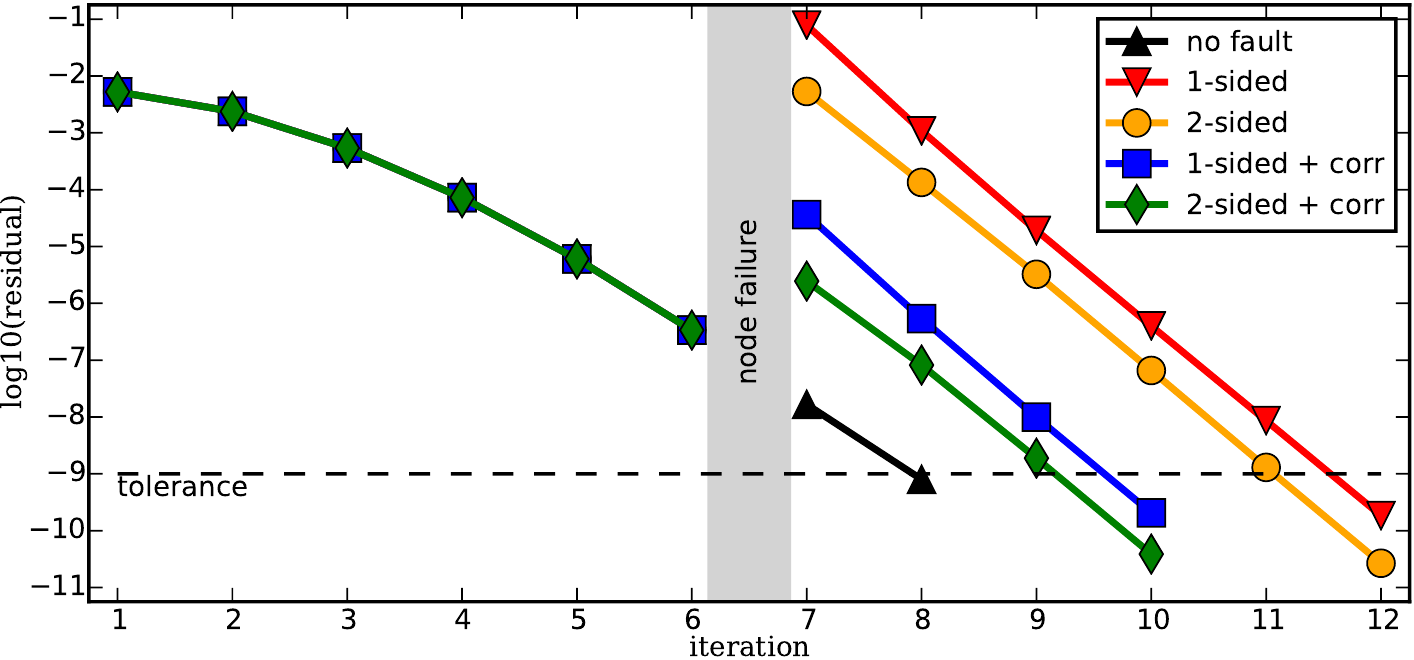}
	\end{minipage}	
	\caption{Residual in time step $7$ against number of iterations with a fault after iteration six for different recovery strategies. The design of this figure is inspired by Figures~4 and~5 in Huber et al.~\cite{RuedeEtAl2015}.}
	\label{fig:residuals}
\end{figure}

Figure~\ref{fig:residuals} shows the PFASST residuals in time step $7$ for both problems plotted against the iteration number $k$.
A fault is simulated in this step after six iterations, i.e. $K_{\text{fault}}=6$.
Before the fault, PFASST is converging well for both problems.
For the heat equation, PFASST reaches the set tolerance of $\varepsilon = 10^{-9}$ after $K=9$ iterations, when no faults are simulated.
Note that for some reason the residual stalls in iteration $7$ and $8$ just above the tolerance and a slightly larger $\varepsilon$ would result in convergence in only $7$ iterations.
The advection equation requires $K=8$ iterations to reach the prescribed tolerance.

The fault after iteration $k=6$ and subsequent recovery increases the residual, since the reconstructed solution is not exact but only approximates lost data.
For both equations, the two-sided reconstruction gives somewhat better results than the one-sided strategy.
However, interpolation alone increases the residual dramatically and causes a massive increase in the number of PFASST iterations required to push the residual below the  tolerance of $10^{-9}$.
For the diffusive problem, simple interpolation requires $K_{\text{add}}=4$ or $K_{\text{add}}=5$ additional iterations, for the advective problem it requires $K_{\text{add}}=3$ or $K_{\text{add}}=4$ additional iterations.
Interpolation alone is clearly not a very efficient strategy.
Note that the residual after the fault is even higher than the one at the beginning of the PFASST run. 
This is due to the weak initial data for the replacement process compared to the initial data used for a standard PFASST run, where a series of coarse sweeps are performed during the predictor phase.

Employing coarse corrections in addition to interpolation significantly reduces the increase of the residual, leading to faster convergence after the fault.
In the cases studied here, the best strategy is two-sided recovery with coarse correction.
It requires only $K_{\text{add}}=1$ additional iterations for the heat equation and $K_{\text{add}}=2$ additional iterations for the advective problem.
Compared to the $K=9$ or $K=8$ iterations in the case without a fault, this corresponds to approximately a $\SI{11}{\percent}$ or $\SI{25}{\percent}$ overhead compared to the no-fault execution plus a small additional overhead from the coarse correction sweeps.
In both cases, this is also much better than the overhead of $K_{\text{fault}}=6$ iterations that a simple restart would incur.

\begin{figure}[!th]
	\centering    
       \begin{minipage}{0.475\textwidth}
        \centering
        \includegraphics[scale=1]{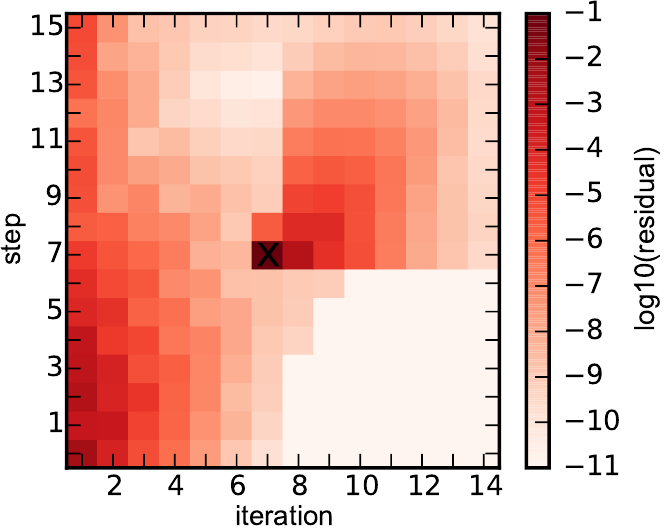}
        \subcaption{1-sided recovery}
        \label{fig:HEAT_steps_vs_iteration_hf_7x7_SPREAD}
    \end{minipage}
    \begin{minipage}{0.475\textwidth}
        \centering
        \includegraphics[scale=1]{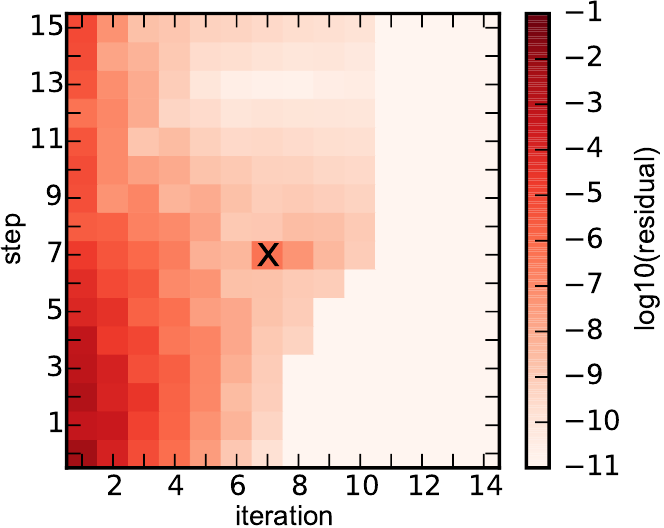}
        \subcaption{2-sided+corr strategy}
        \label{fig:HEAT_steps_vs_iteration_hf_7x7_INTERP_PREDICT}
    \end{minipage}
    \caption{Example with fault injection at step (i.e. processor) 7, iteration 7 and impact on residuals for the heat equation, see~\eqref{eq:hard_prob_A}.
    \label{fig:hard_heat_stepsvsiter}}
\end{figure}

These results show only the effect of a single fault and subsequent recovery on the residual of a single time step.
Since each time step communicates information to its successors, the increase in the residual will propagate forward in time and affect other time steps.
This is illustrated in Figure~\ref{fig:hard_heat_stepsvsiter}, which shows the residual at all time steps (y-axis) in all iterations (x-axis) for two different recovery strategies.
In the no-fault case (not shown), PFASST requires $K=9$ iterations to converge.
For the simple one-sided recovery shown in Figure~\ref{fig:HEAT_steps_vs_iteration_hf_7x7_SPREAD}, the increased residual after a fault in step $7$ after iteration $6$ not only affects subsequent iterations on time step $7$ but spreads to later time steps and pollutes their solutions, too.
Using a better strategy leads to much smaller impact on later time steps as shown in Figure~\ref{fig:HEAT_steps_vs_iteration_hf_7x7_INTERP_PREDICT}.
Even though there is still a small increase in residual in later time steps after the fault, time steps $7$ to $15$ converge quickly within a total of $K=10$ iterations, i.e. $K_\text{add}=1$.

\begin{figure}[!th]
	\centering    
    \begin{minipage}{0.475\textwidth}
        \centering
        \includegraphics[scale=1]{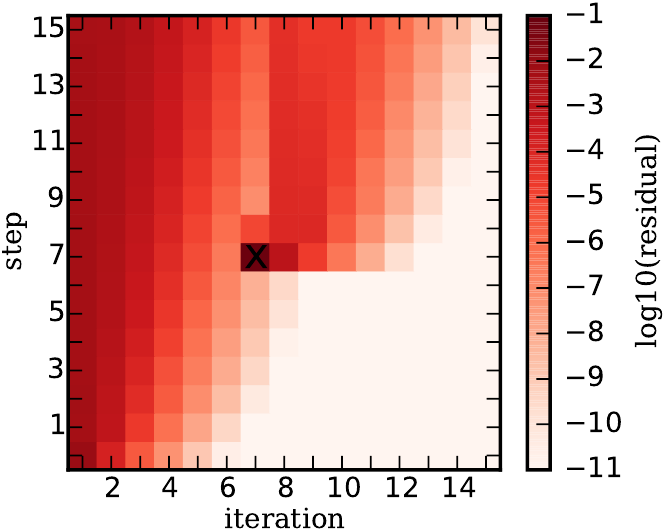}
        \subcaption{1-sided recovery}\label{fig:ADVECTION_steps_vs_iteration_hf_7x7_SPREAD}
    \end{minipage}
    \begin{minipage}{0.475\textwidth}
        \centering
        \includegraphics[scale=1]{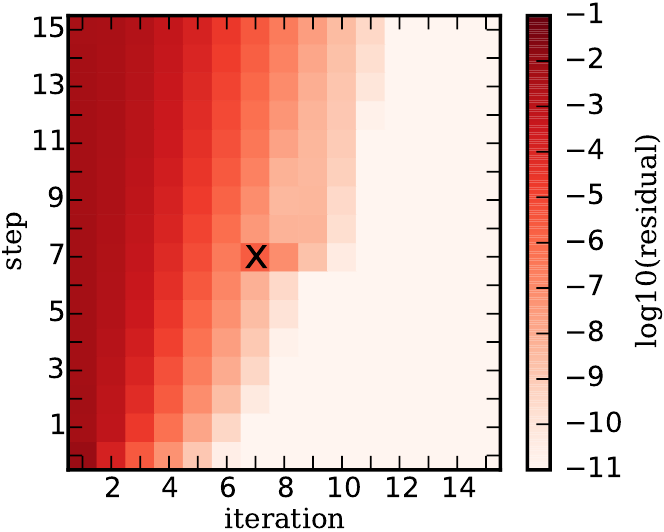}
        \subcaption{2-sided+corr recovery}\label{fig:ADVECTION_steps_vs_iteration_hf_7x7_INTERP_PREDICT}
    \end{minipage}
    \caption{Example with fault injection at step (i.e. processor) 7, iteration 7 and impact on residuals for the advection equation, see~\eqref{eq:hard_prob_B}.
    \label{fig:hard_adv_stepsvsiter}}
\end{figure}

Essentially the same behavior is seen for the advection equation in Figure~\ref{fig:hard_adv_stepsvsiter}.
Here, the no-fault run takes $K=11$ iterations to converge. 
Because there is no diffusion, pollution of later time steps from incomplete recovery is worse than for the diffusive problem. 
For the one-sided interpolation strategy depicted in Figure~\ref{fig:ADVECTION_steps_vs_iteration_hf_7x7_SPREAD}, the fault in time step $7$ pollutes all later time steps in the next iteration equally, increasing their residual by several orders of magnitude.
In contrast, two-sided interpolation with coarse corrections shown in Figure~\ref{fig:HEAT_steps_vs_iteration_hf_7x7_INTERP_PREDICT} only marginally affects later time steps and allows for convergence of all later steps within $4$ iterations, which results in a perfect $K_\text{add} = 0$.
 
\begin{figure}[t!]

    \begin{minipage}{0.475\textwidth}
        \centering
        \includegraphics[scale=1]{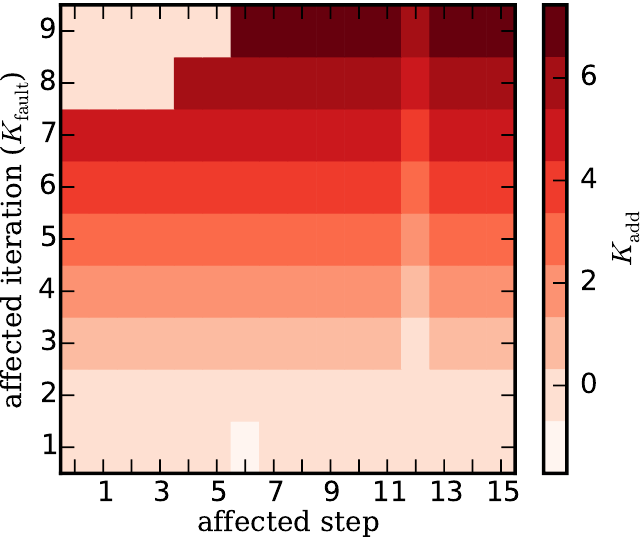}
        \subcaption{1-sided recovery}\label{fig:HEAT_iteration_counts_hf_SPREAD}
    \end{minipage}
    \begin{minipage}{0.475\textwidth}
        \centering
        \includegraphics[scale=1]{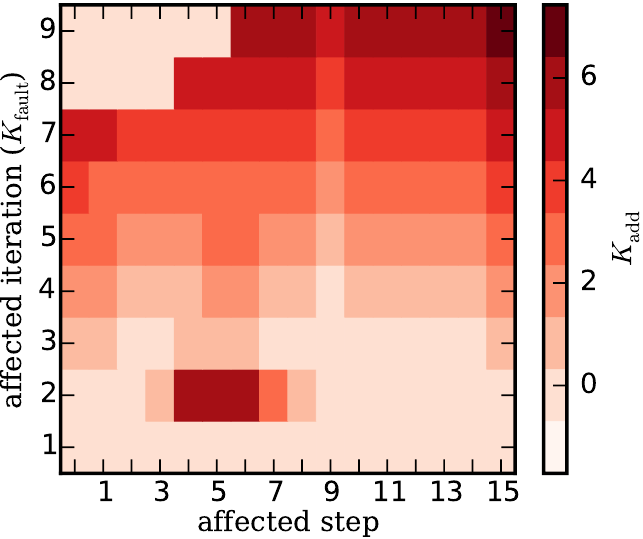}
        \subcaption{2-sided recovery}\label{fig:HEAT_iteration_counts_hf_INTERP}
    \end{minipage}\\[\baselineskip]
    \begin{minipage}{0.475\textwidth}
        \centering
        \includegraphics[scale=1]{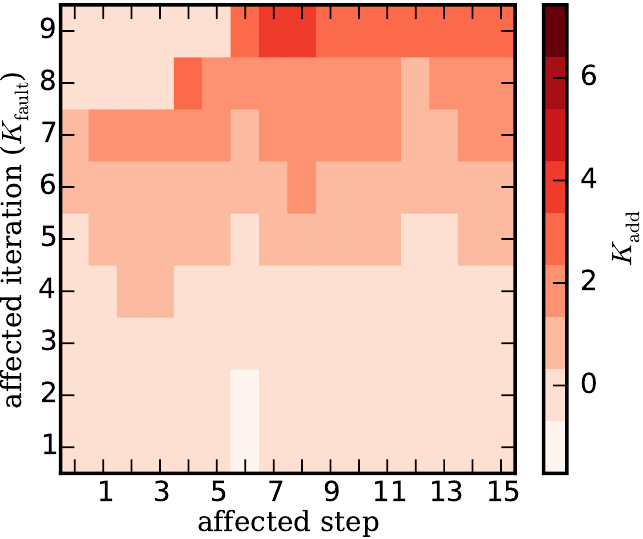}
        \subcaption{1-sided+corr recovery}\label{fig:HEAT_iteration_counts_hf_SPREAD_PREDICT}
    \end{minipage}
    \begin{minipage}{0.475\textwidth}
        \centering
        \includegraphics[scale=1]{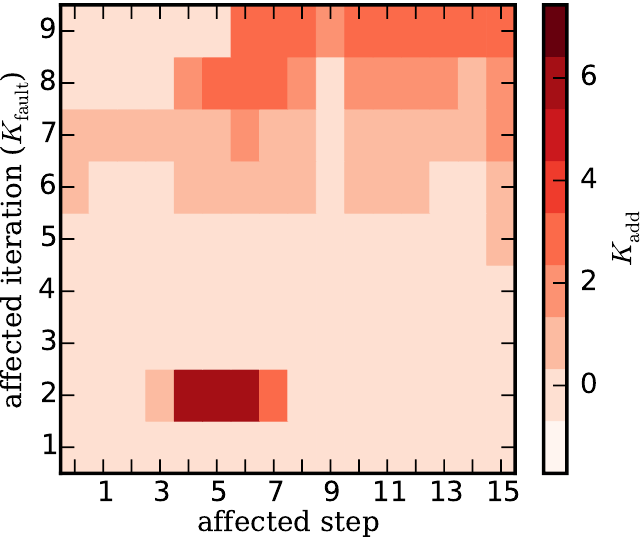}
        \subcaption{2-sided+corr recovery}\label{fig:HEAT_iteration_counts_hf_INTERP_PREDICT}
    \end{minipage}
    \caption{Number of additional iterations $K_\text{add}$ for the heat equation~\eqref{eq:hard_prob_A} if a fault occurs at a particular step ($x$ axis) and a particular iteration ($y$ axis).
    \label{fig:hard_heat_iterheatmap}\\[2\baselineskip]}
\end{figure}

\begin{figure}[t!]
    \begin{minipage}{0.475\textwidth}
        \centering
        \includegraphics[scale=1]{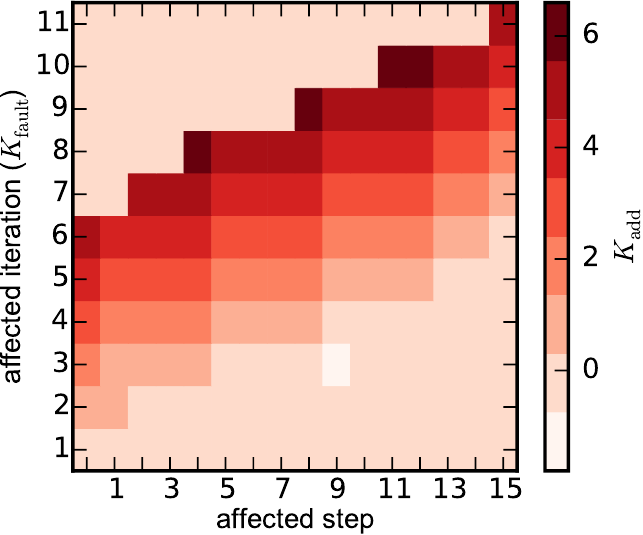}
        \subcaption{1-sided recovery}\label{fig:ADVECTION_iteration_counts_hf_SPREAD}
    \end{minipage}
    \begin{minipage}{0.475\textwidth}
        \centering
        \includegraphics[scale=1]{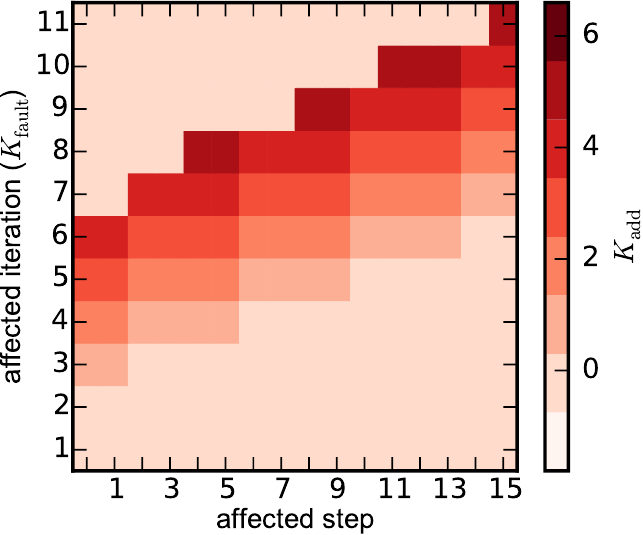}
        \subcaption{2-sided recovery}\label{fig:ADVECTION_iteration_counts_hf_INTERP}
    \end{minipage}\\[\baselineskip]
    \begin{minipage}{0.475\textwidth}
        \centering
        \includegraphics[scale=1]{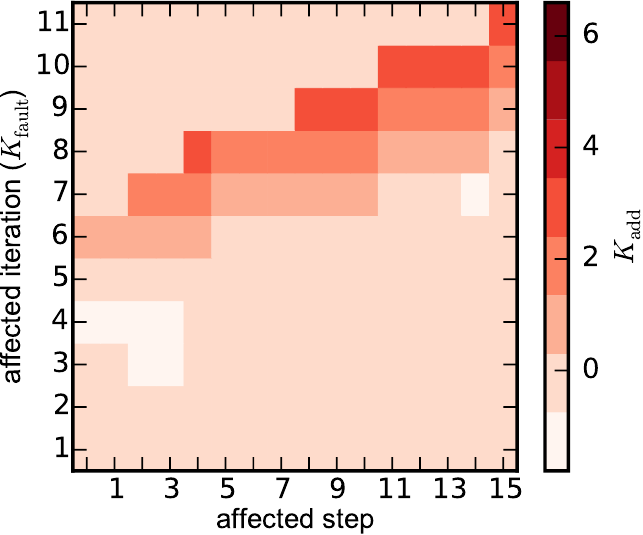}
        \subcaption{1-sided+corr recovery}\label{fig:ADVECTION_iteration_counts_hf_SPREAD_PREDICT}
    \end{minipage}
    \begin{minipage}{0.475\textwidth}
        \centering
        \includegraphics[scale=1]{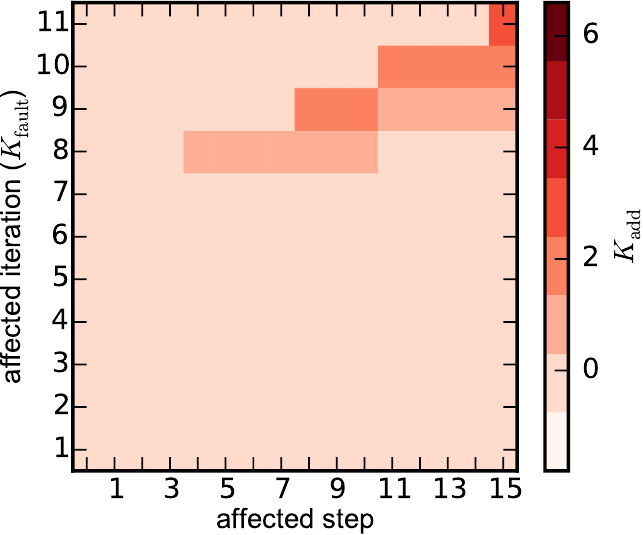}
        \subcaption{2-sided+corr recovery}\label{fig:ADVECTION_iteration_counts_hf_INTERP_PREDICT}
    \end{minipage}
    \caption{Number of additional iterations $K_\text{add}$ for the advection equation~\eqref{eq:hard_prob_B} if fault occurs at a particular step ($x$ axis) and a particular iteration ($y$ axis).
    \label{fig:hard_adv_iterheatmap}\\[2\baselineskip]}
\end{figure}

So far, a fault was only simulated at one specific time step in one specific iteration.
However, both time step as well as iteration number where a fault occurs will have an important influence on how the fault affects convergence.
For both setups, Figures~\ref{fig:hard_heat_iterheatmap} and~\ref{fig:hard_adv_iterheatmap} show parameter studies for the four different strategies, where both the affected iteration $K_\text{fault}$ (y-axis) and time step (x-axis) are varied.
The total number of required iterations is color-coded with larger numbers being darker.

For the diffusive problem, the no-fault case takes $K=9$ iterations. 
In Figure~\ref{fig:hard_heat_iterheatmap}, the impact of the four recovery strategies can be seen. 
Darker colors indicate higher numbers of additional iterations $K_\text{add}$, which vary from $-1$ to $7$.
Besides the small regions, where up to $6$ additional iterations are required at $K_\text{fault} = 2$, the two-sided strategy with additional coarse correction is clearly superior, requiring only $K_\text{add} = 3$ additional iterations.
The one-sided strategy with coarse correction is slightly less effective.
When using recovery without coarse correction, the number of additional iterations mainly depends on the affected iteration, while the affected step plays only a minor role.
In general, the later the affected iteration, the larger $K_\text{add}$ becomes.
The reason is that in later iterations, the approximate solution is already very accurate so that the error made during recovery has a more significant impact.
In early iterations, the solution is still so inaccurate that the recovery error has almost no effect.
Most importantly, in all cases we have $K_\text{add}<K_\text{fault}$ so that the recovery strategies produce less overhead than a hard restart.

For the advective problem the no-fault case takes $K=11$ iterations.
Figure~\ref{fig:hard_adv_iterheatmap} shows the impact of the four recovery strategies in this case.
Again, darker colours indicate higher numbers of additional iterations $K_\text{add}$, which vary from $-1$ to $6$.
Both one-sided and two-sided interpolation with coarse correction are effective strategies.
The latter can recover faults up to $K_{\text{fault}}=7$ without any additional iterations.
Even a fault in the last iteration in the last time step, which is the worst-case, only causes $K_{\text{add}}=3$ additional iterations compared to the overhead of $K_{\text{fault}}=11$ for a simple restart.
In contrast to the heat equation, faults in later time steps cause fewer additional iterations, unless they occur in late iterations.
Again, all strategies satisfy the criterion $K_\text{add}<K_\text{fault}$ for efficiency compared to restarting.

In some cases we observe one of two rather odd situations: (1) a recovery after an early fault (i.e.~small $K_\text{fault}$) can actually lead to a reduction of iterations compared to the no-fault case and (2) sometime recovery strategies can lead to a massive increase in iterations while performing well for most other cases.
Incident (1) can be observed e.g. in Figure~\ref{fig:HEAT_iteration_counts_hf_SPREAD}, step 6, iteration 1 or in Figure~\ref{fig:ADVECTION_iteration_counts_hf_SPREAD_PREDICT}, steps 0 to 3, iterations 3 to 4.
It seems that the recovery strategies provide slightly better initial data in this early phase, so that the algorithm converges faster here, saving one iteration.
In our experiments, incident (2) seems to occur only for the heat equation example when using the two-sided recovery strategy, i.e.~in Figures~\ref{fig:HEAT_iteration_counts_hf_INTERP} and~\ref{fig:HEAT_iteration_counts_hf_INTERP_PREDICT}. 
When using a higher resolution in space (511 degrees-of-freedom instead of 255) and thus a better coarse level resolution, this effect does not occur. 
In turn, using less degrees-of-freedom and thus a worse coarse level resolution results in extremely slow convergence even when no fault occurs. 
Therefore, the quality of the coarse level is crucial not only for convergence of the PFASST algorithm itself  (which is well-known) but also for the effectiveness of recovery strategies.

%

\subsection{Gray-Scott reaction-diffusion}\label{subsec:gray-scott}
As a more complex diffusive example we consider the 1D Gray-Scott model~\cite{Gray1983-im} for a chemical reaction of two components $\mathcal{U}$ and $\mathcal{V}$.
The model is given by the reaction-diffusion equations
\begin{equation}
    \label{eq:grayscott}
    \begin{aligned}
        u_t &= \Delta u - uv^2 + A(1-u),\\
        v_t &= D\Delta v + uv^2 - Bu,
    \end{aligned}
\end{equation}
where $u = u(x,t)$ and $v = v(x,t)$ are the concentrations of the two species $\mathcal{U}$ and $\mathcal{V}$, $D$ is the normalized diffusion coefficient of $\mathcal{V}$, $A$ denotes the fed rate into the system (e.g.~a reactor) and $B$ is the overall decay rate of $\mathcal{V}$.
Using this model, chemical reactions of the type $\mathcal{U} + 2\mathcal{V} \rightarrow 3\mathcal{V}$, $\mathcal{V}\rightarrow\mathcal{P}$ can be simulated, where $\mathcal{P}$ is some inert product of the reaction.
Investigation of this model and its dynamics is an active topic of research~\cite{Doelman1997-zu,Doelman1998-sd,Reynolds1994-qd}.

\begin{figure}[!t]
    \begin{minipage}{0.32\textwidth}
        \centering
        \includegraphics[width=\textwidth]{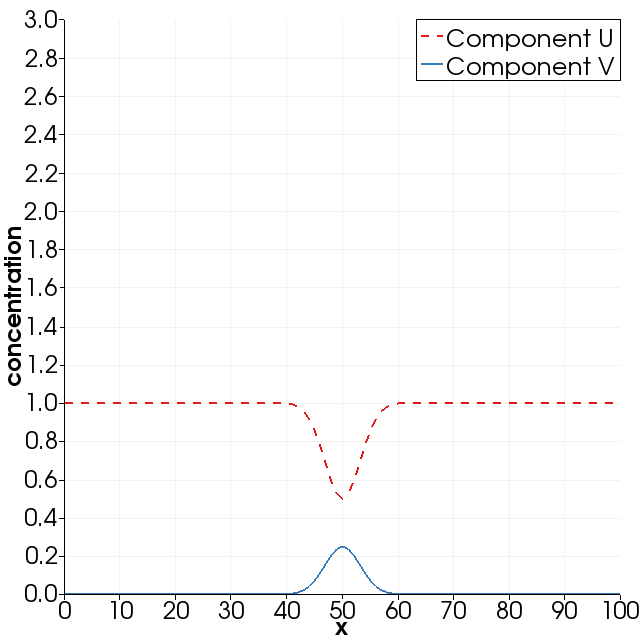}
        \subcaption{Initial concentrations}\label{fig:grayscott_step00}
    \end{minipage}
    \begin{minipage}{0.32\textwidth}
        \centering
        \includegraphics[width=\textwidth]{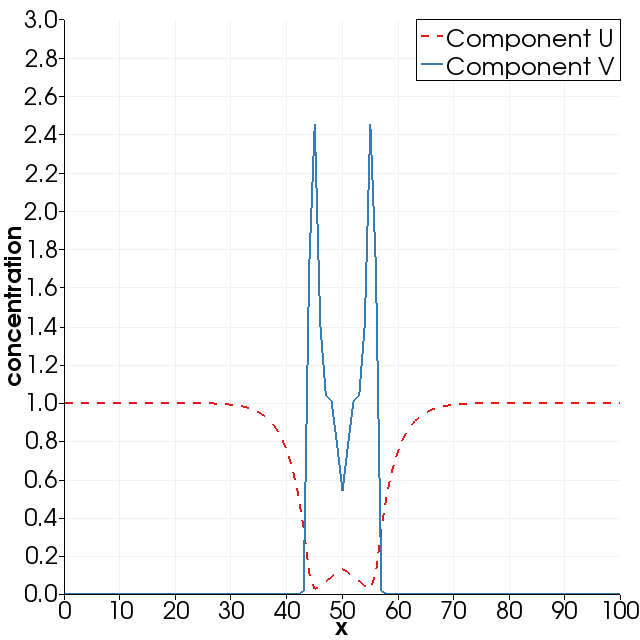}
        \subcaption{Time $t=100$}\label{fig:grayscott_step100}
    \end{minipage}
    \begin{minipage}{0.32\textwidth}
        \centering
        \includegraphics[width=\textwidth]{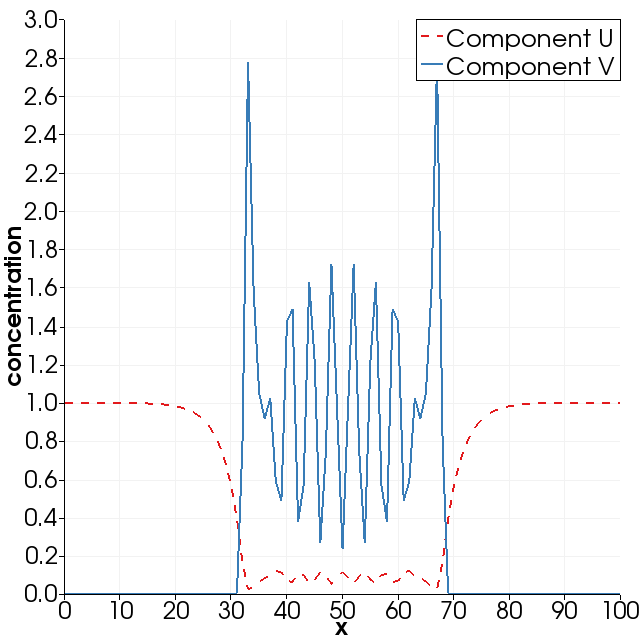}
        \subcaption{Time $t=400$}\label{fig:grayscott_step400}
    \end{minipage}
    \caption{Evolution of the concentrations of components $\mathcal{U}$ and $\mathcal{V}$ over time.\label{fig:grayscott_evolution}}  
\end{figure}

We select $A=0.09$, $B=0.086$ and $D=0.01$, which corresponds to the setup used to generate Figure~10 in Doelman et al.~\cite{Doelman1997-zu} and leads to a dynamical evolution of pulses as shown in Figure~\ref{fig:grayscott_evolution}. 
We start with initial conditions
\begin{equation}
    u(x,0) = 1 - \frac{1}{2}\sin^{100}(\pi x/L),\quad v(x,0) = \frac{1}{4}\sin^{100}(\pi x/L)
\end{equation}
for $x\in[0,L]=[0,100]$ in our case, representing a sharp initial peak at the center of the domain, see Figure~\ref{fig:grayscott_step00}.
We use homogeneous Neumann boundary conditions.

For the spatial discretization, we use the FEniCS framework~\cite{Logg2012-lo} and in particular the user interface DOLFIN~\cite{Logg2010-em} with its Python front-end.
For this project, we extended pySDC to handle FEniCS' weak formulation of PDEs and it is now capable of handling complicated multi-component equations by exploiting FEniCS' formalism.
By specifying the right-hand sides in weak form and using FEniCS's built-in solvers, pySDC provides easy-to-use high-order time-stepping for finite element discretizations.

We choose $3$ Gauss-Radau collocation nodes (with right end-point included) for fifth-order accuracy in time. 
We also use a LU decomposition of the quadrature matrix to speed up SDC's and PFASST's convergence~\cite{Weiser2014}\footnote{Colloquially, this strategy has become known as ''St.~Martin's trick'' due to the first name of its inventor.}.
The simulation is run until to $T=1280.0$ with time step $\Delta t = 2.0$, i.e.~for $640$ time steps.
To properly resolve the sharp pulses in space, we use fourth-order standard finite elements with $N=513$ degrees-of-freedom.
FEniCS' built-in Newton method serves as spatial solver with absolute tolerance $10^{-9}$ and relative tolerance $10^{-8}$, treating the full right-hand side of the PDE implicitly.
We parallelize in time using $20$ blocks of $32$ parallel steps with two-level PFASST. 
Standard coarsening via reduction of degrees-of-freedom is employed, i.e. we use $N=257$ degrees-of-freedom on the coarse level.
PFASST iterates until a residual of $10^{-7}$ is reached (absolute tolerance).
\begin{figure}[!t]
    \centering
    \includegraphics[scale=1]{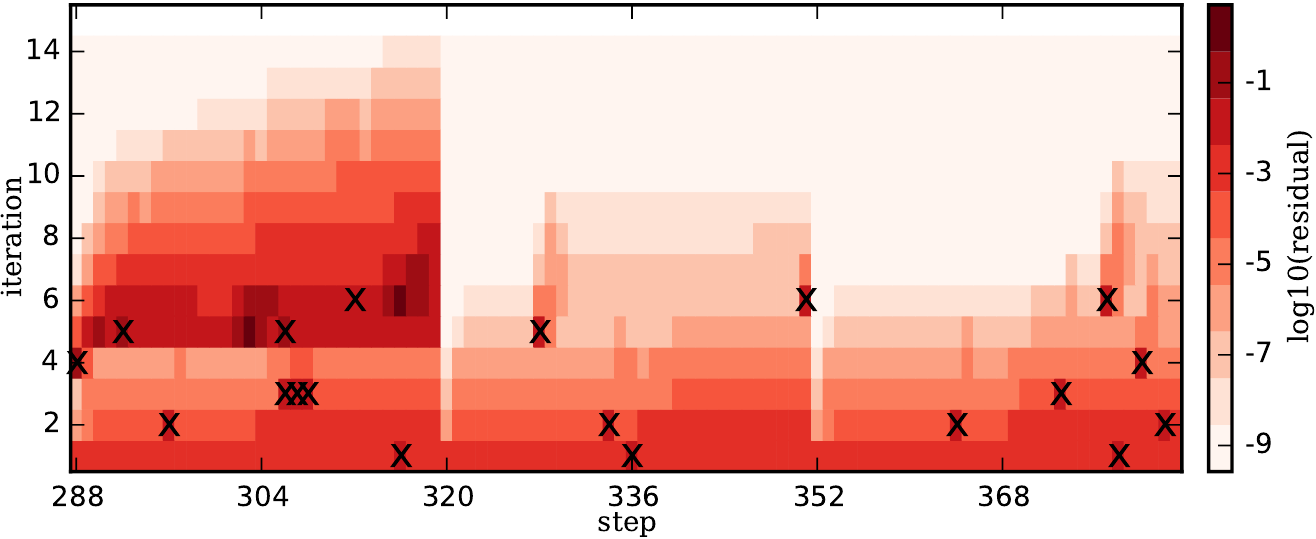}
    \caption{Evolution of the residual for the Gray-Scott example~\eqref{eq:grayscott} using $32$ parallel steps per block, showing $3$ representative blocks out of $20$ (blocks $9$ to $11$), which comprise the steps $288$ to $384$ and cover the time interval $t=576$ to $t=768$. Failures are marked with an \textbf{x}. Simple one-sided recovery without coarse correction sweeps is performed.}
    \label{fig:GRAYSCOTT_steps_vs_iteration_hf_SPREAD}
\end{figure}
\begin{figure}[!th]
    \centering
    \includegraphics[scale=1]{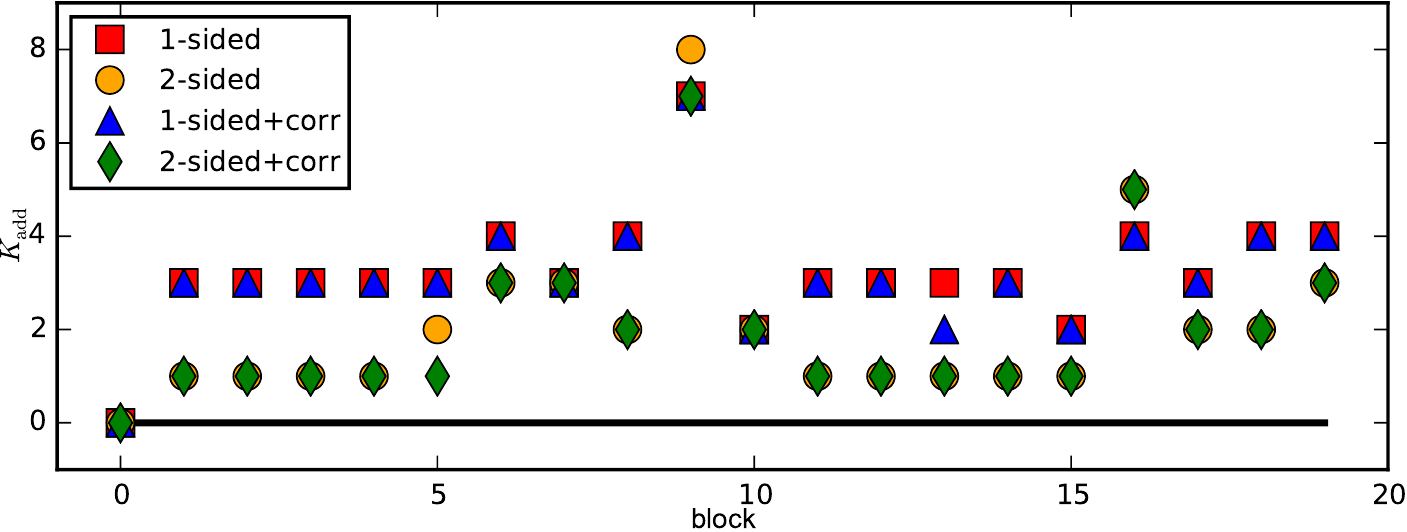}
    \caption{Additional iterations $K_{\text{add}}$ for each block in the Gray-Scott example using the four different recovery strategies. We only show iteration counts of the last process for each block to avoid cluttering. The black line indicates the the base line of the run without faults.
    \label{fig:GRAYSCOTT_Kadd_vs_NOFAULT_hf}} 
\end{figure}

For the no-fault run, the maximum number of iterations is $K=7$ except for the first block which needs $K=10$ iterations due to its fast dynamics. 
To ``stress-test'' the different recovery strategies, faults are injected at random. 
Before starting a new iteration, we inject a fault with a probability of $3\%$.
This probability is clearly very high and (hopefully) does not reflect hardware properties of real-world HPC systems. 
It however creates different realistic failure patterns throughout the run, which are difficult to anticipate and create a-priori.
It also minimizes the chance of testing the recovery strategies for favorable conditions only.
Interesting structures observed in the distribution of faults are e.g.~multiple faults of the same process in quick succession, which, in a real-world system, could be caused by a faulty component or clusters of faults mimicking cascading failures.
In order to be able to compare different strategies, the random pattern of faults is generated a-priori and then applied to all simulations testing different recovery strategies.
We only allow faults in iterations that are also performed in the no-fault case, that is additional iterations caused by fault recovery are not subjected to faults.
Otherwise, comparison with the no-fault reference becomes difficult.

The simplest recovery strategy is applied, one-sided interpolation without coarse level corrections.
In Figure~\ref{fig:GRAYSCOTT_steps_vs_iteration_hf_SPREAD}, three representative time-parallel blocks are shown, ranging from $t=578$ to $t=768$ (blocks $9$, $10$, $11$) which contain time steps $288$ to $384$.
In particular, this part of the run contains the worst result using one-sided recovery: up to $K_\text{add}=7$ additional iterations are required to converge in block $9$ (steps $288$--$320$).
This is due to the cluster of failures in iterations $3$ and $5$. 
Darker colors indicate higher residuals on this process and while the two isolated faults in iterations $1$ and $2$ have negligible impact on convergence, the three failures in iteration $3$ as well as the failures in iterations $5$ and $6$ lead to very high residuals on all subsequent processes.
In the next block, only four faults are injected and their impact is limited, leading to $K_\text{add}=2$ additional iterations which are mainly due to the isolated faults in iterations $5$ and $6$.
The last bock shows a cluster of faults: during six iterations there are four faults within the last seven processes. 
While the previous processes converge rapidly within less than the original $K=7$ iterations, this last group of affected processes needs another $K_\text{add}=3$ iterations to finish.
We emphasize, however, that even the most simple recovery strategy performs quite well:
except for the first block, we always have $K_\text{add}<K_\text{fault}$ despite multiple failures.
Since simple restarting would have to restart after every single fault and probably not make any progress at all, recovery is clearly the more efficient option.

In Figure~\ref{fig:GRAYSCOTT_Kadd_vs_NOFAULT_hf} we show the additional iterations $K_\text{add}$ caused by the faults for the four recovery strategies.
For clarity and in contrast to Figure~\ref{fig:GRAYSCOTT_steps_vs_iteration_hf_SPREAD}, we only consider the iteration count at the last process of each block, but now show data for the whole run containing all $20$ blocks. 
Since the last process in each block is always the last one to converge, its iteration count determines the iteration count of the whole PFASST block.
Except for blocks $0$, $9$ and $16$, using two-sided interpolation is always better than one-sided interpolation.
In all other blocks, the two-sided strategy is clearly superior, reducing $K_\text{add}$ by one or two.

Interestingly, applying coarse corrections does not improve performance further, neither for one- nor for two-sided interpolation.  
Only at blocks $5$ and $9$ does the coarse correction reduce $K_\text{add}$ compared to solely two-sided interpolation.
Two-sided interpolation alone already provides an effective recovery strategy and the more expensive application of coarse corrections is, for this configuration, not justified.
This is related to the observation made before: the quality of the coarse level determines not only the convergence of the PFASST algorithm itself but also the effectiveness of recovery strategies.
Tests with twice as many degrees-of-freedom show that when overall spatial resolution of the problem is higher, coarse corrections do help to reduce $K_\text{add}$ further.

\subsection{Boussinesq equations}\label{subsec:boussinesq}
As complex benchmark of hyperbolic type we consider the linearized Boussinesq equations~\cite[Section 8.2.4]{Durran2010}
\begin{equation}
	\label{eq:boussinesq}
	\begin{aligned}
		u_t + U u_x + c_s p_x &= 0, \\
		w_t + U w_x + c_s p_z &= b, \\
		b_t + U b_x + N^2 w &= 0, \\
		p_t + U p_x + c_s \left( u_x + w_z \right) &= 0,
	\end{aligned}
\end{equation}
a transformed variant of the Euler equations.
Equations~\eqref{eq:boussinesq} describe flow of a compressible, stably stratified fluid.
Here, $(u,w)$ is the velocity field, $b$ the buyoancy and $p$ the pressure.
The parameter $U$ is the advection velocity of the background, $c_s$ is the acoustic wave speed while $N$ is a parameter governing the stability of the stratification of the fluid, the so-called Brunt-V\"ais\"al\"a frequency.
We consider a test case widely used in meteorology, where an initial perturbation in buoyancy generates a gravity wave traveling through a channel~\cite{WickerSkamarock1998}.

Equations~\eqref{eq:boussinesq} are discretized in space on the fine level using fifth order upwind stencils for the advective terms and fourth order centered differences for the other derivatives.
On the coarse level, first order upwind stencils together with second order centered stencils are used.
Only operator coarsening is used here, but no reduction of degrees-of-freedom on the coarse levels.
The computational domain is $[-150, 150] \times [0, 10]$, corresponding to a channel with a length of \SI{300}{\kilo\metre} and a height of \SI{10}{\kilo\metre}.
Both levels in PFASST use \SI{450}{} nodes in the horizontal and \SI{30}{} nodes in the vertical direction, resulting in a slightly anisotropic mesh where the vertical resolution is twice as fine as the horizontal resolution.
The horizontal background velocity is set to $U=0.02$, corresponding to \SI{20}{\metre\per\second}, the acoustic wave velocity to $c_s=0.3$, corresponding to \SI{300}{\metre\per\second}. 
Both are realistic values for wind and sound speeds in the troposphere.
The stability parameter is set to $N=\SI{0.01}{\per\second}$.
Periodic boundary conditions in horizontal direction and a no-slip condition at the top and bottom are used.

\begin{figure}[!thp]
	\centering
	\includegraphics[scale=1]{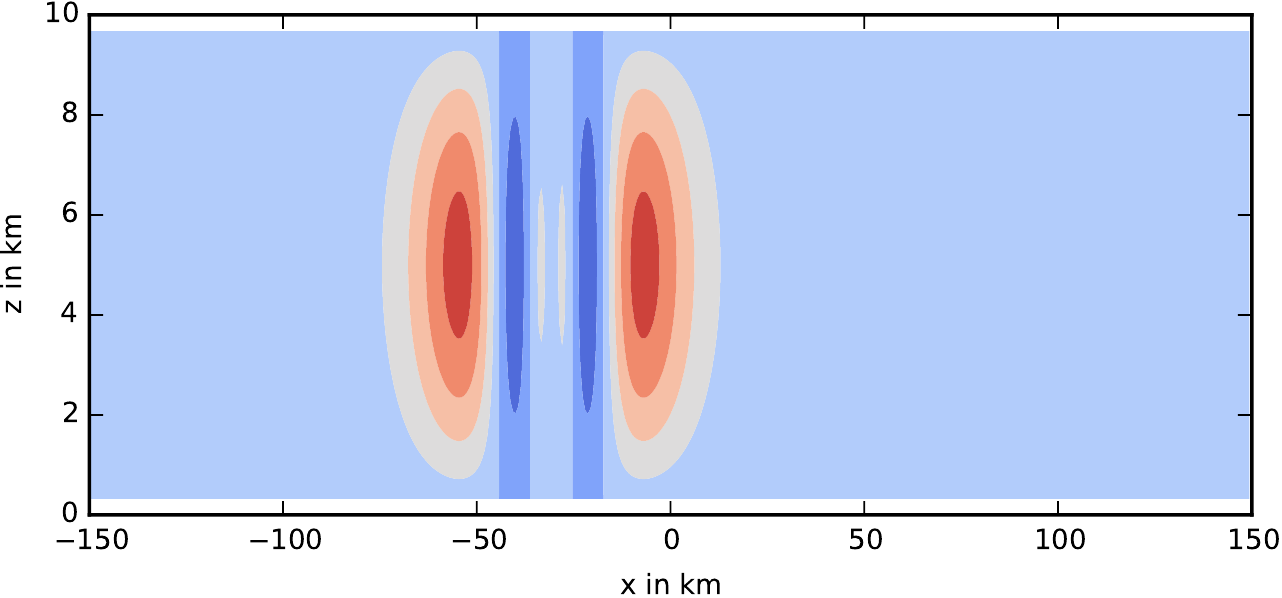}
	\caption{Buoyancy $b$ at $t=\SI{960}{\second}$ for the Boussinesq equations. The initial perturbation has created gravity waves travelling to the left and right in the channel while background advection moves them slowly to the right. The difference between isolines is $0.001$ with negative values in dark blue and positive values in red.}
	\label{fig:boussinesq}
\end{figure}

PFASST integrates the problem until $T=\SI{960}{\second}$ on $P=16$ processors with $M=3$ Gauss-Lobatto nodes per time step and a tolerance of $\varepsilon=10^{-6}$.
The time step size is $\Delta t = \SI{3}{\second}$, leading to $320$ time steps and thus $20$ PFASST blocks.
A SDC sweep with fast-wave slow-wave splitting~\cite{RuprechtSpeck2016} is used, where the slow advective terms are integrated explicitly, while the terms associated with fast traveling gravity and acoustic waves are integrated implicitly.
To solve the implicit part, GMRES with a tolerance of $10^{-10}$ and restarting after ten iterations is used.
A localized perturbation of buoyancy 
\begin{equation}
	\theta(x,z,0) = \Delta \theta \sin\left( \frac{\pi z}{H} \right)\frac{1}{1 + \frac{\left(x-x_c\right)^2}{a^2}}
\end{equation}
with $x_c = -50$, $a=5$, $H=10$ and $\Delta \theta = 0.01$ is set as initial data while $u$, $w$ and $p$ are initially set to zero.
This initial buoyancy perturbation generates gravity waves travelling to the left and to the right through the channel, see Figure~\ref{fig:boussinesq}.
Since the model is compressible, sounds waves are present as well but too small in amplitude to be visible in the depicted solution.

\begin{figure}[!th]
    \centering
    \includegraphics[scale=1]{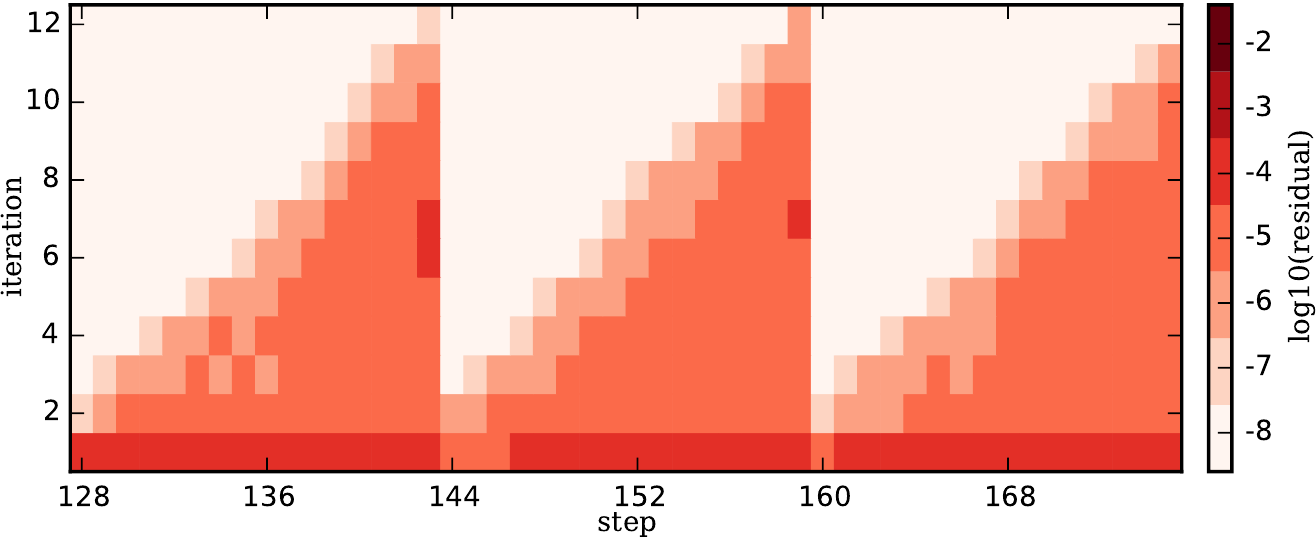}
    \caption{Evolution of the residual for the Boussinesq example~\eqref{eq:boussinesq} without faults. 
Shown are the $3$ representative blocks $9$, $10$ and $11$ out of $20$, which corresponds to time steps $128$ to $176$ and cover the time interval $[384,528]$. \label{fig:BOUSSINESQ_steps_vs_iteration_hf_NOFAULT}\\}
    
    \centering 
    \includegraphics[scale=1]{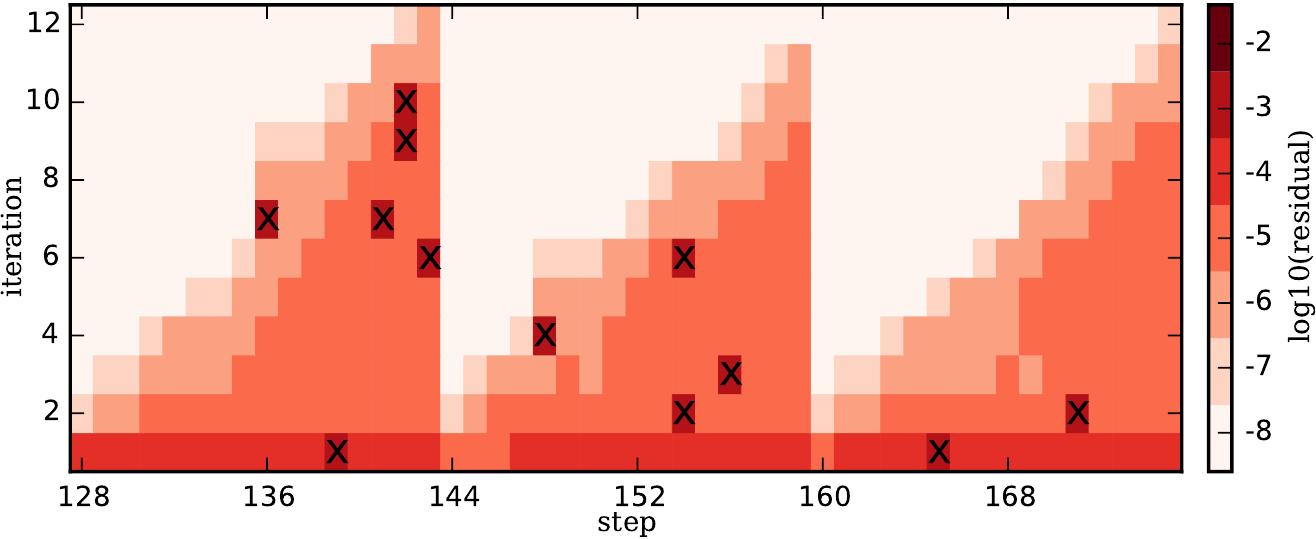}
    \caption{Evolution of the residual for the Boussinesq example~\eqref{eq:boussinesq} with faults.
Shown are the $3$ representative blocks $9$, $10$ and $11$ out of $20$, which correspond to time steps $128$ to $176$ and cover the time interval $[384,528]$.
Simulated hardware failures are marked by an \textbf{x}. 
Only one-sided recovery without coarse correction sweeps is used. \label{fig:BOUSSINESQ_steps_vs_iteration_hf_SPREAD}\\}    
\end{figure}

Figure~\ref{fig:BOUSSINESQ_steps_vs_iteration_hf_NOFAULT} shows the evolution of the PFASST residual for the Boussinesq equation for three consecutive blocks of sixteen steps without faults.
Shown are the ninth, tenth and eleventh blocks, covering the time from $t=\SI{384}{\second}$ to $t = \SI{528}{\second}$.
Other blocks show essentially the same behavior.
As can be seen, for the hyperbolic example PFASST converges noticeably slower than for the diffusive Gray-Scott problem.
In all blocks, the last processor needs between $K=11$ and $K=12$ iterations to reach the requested tolerance.
The step-shaped structure indicates that time steps converge at a rate of about one per iteration, in contrast to the Gray-Scott example where usually multiple processors converge in the same iteration.

Figure~\ref{fig:BOUSSINESQ_steps_vs_iteration_hf_SPREAD} shows the same blocks but now with randomly injected faults, indicated by black crosses.
Only simple one-sided recovery without coarse corrections is performed here.
For the Boussinesq problem, all four analyzed recovery strategies perform essentially the same. 
Thus we show results only for the most simple one, since the higher cost of the more elaborate procedures seem to provide no benefit for this particular setup.
Again, using better resolution in space may lead to more pronounced differences between the recovery strategies. 

The impact of the faults on the number of iterations is small: the last block needs one additional iteration ($K_{\text{add}}=1$) to converge while, for some reason, the middle block requires one iteration less in the case with fault recovery. 
In this particular case, the fault removes an unfavorable value from the iteration, similar to the incidents described in Section~\ref{subsec:simple_problems}.
Faults still do have a visible impact on the evolution of the residual, see e.g. the fault in iteration 7 in step 136 which causes a small localized increase in residual in later iterations in subsequent time steps.
However, these disturbances have only a small effect on the convergence of the last processor, which determines overall cost.
Therefore, even with a rather basic recovery strategy, fault-tolerant PFASST allows the simulation to progress at almost the same rate as in the no-fault case.

\section{Outlook and possibilities for further research}\label{sect:outro}

\paragraph{Extension to multi-level PFASST} One direction for further research is to study the impact on fault recovery when using multiple levels in PFASST.
By design, PFASST can be used with a full space-time hierarchy.
Application of the recovery strategies presented should lead to lower recovery costs when more than two levels are accessible.
However, the influence on the number of additional iterations $K_\text{add}$ is not obvious and detailed tests are required to evaluate the final overhead. 
There may also exist more involved strategies for recovery, especially when PFASST runs with more than two levels.
Also, it is worth investigating whether or not processes waiting for the recovery procedure to finish can do other meaningful work, e.g.~perform additional iterations as in the predictor phase.
This could potentially reduce $K_\text{add}$ further.

\paragraph{Extension of the overhead model} The convergence of the spatial iterative solver, e.g. the Newton solver for the Gray-Scott example in Section~\ref{subsec:gray-scott} or GMRES for the Boussinesq equation in Section~\ref{subsec:boussinesq}, benefit from recovery strategies as well. 
As PFASST converges, the accuracy of the initial solutions provided for the spatial solver improves, which means that in later iterations they converge quicker and implicit solves get much cheaper~\cite{MinionEtAl2015,SpeckEtAl2016}.
With reasonably accurate solutions after the recovery attempt, the spatial solver starts with better initial values on the fine level.
For a complete restart, in contrast, these accurate initial values are not available which will lead to further increase of overhead.
Since the model in Section~\ref{subsec:overhead} assumes constant costs for coarse or fine sweeps, this effect is not covered.
A more detailed investigation is left for future work.

\paragraph{Combination with parallelization in space}
In most cases, parallel-in-time integration will be used in combination with large-scale spatial parallelization.
Since spatial parallelization usually employs some form of decomposition of the spatial domain, a single failing process does not mean that the full solution at a time step is lost but only the solution of one spatial subdomain.
The straightforward generalization of the here presented strategies would be a local recovery, using only information from the solution on the same subdomain at the time step before and after.
A direct continuation of this work would be to assess the impact of the different recovery strategies in such a scenario.
Since approaches for algorithm-based fault tolerance and fault recovery for spatial solvers already exist, more intricate forms of combined space-time recovery procedures could be devised as well.

\paragraph{Generalization to other parallel-in-time integration methods}
The fundamental idea of the algorithm-based recovery approach presented here relies on the iterative nature of PFASST.
It is natural to transfer the concept to other iterative parallel-in-time integration schemes such as Parareal~\cite{LionsEtAl2001} or MGRIT~\cite{FalgoutEtAl2014_MGRIT}.
In these methods, each level (two for Parareal, multiple for MGRIT) is equipped with standard, non-iterative propagators such as implicit Euler or Runge-Kutta methods.
Using similar strategies as presented here, it seems likely that these method lend themselves equally well to algorithm-based fault tolerance.
Investigating recovery strategies in the context of Parareal and/or MGRIT is an interesting topic for further studies.

\paragraph{Implementation into an MPI framework}
While the paper illustrates that the PFASST algorithm is a promising candidate for algorithm-based fault tolerance, the feasibility of the approach in large-scale parallel computations remains to be shown.
An important next step would be to validate that the theoretical model of overhead in terms of iterations gives an accurate representation of overhead in terms of wall clock time.
For the Python framework pySDC, implementing fault injections and testing new strategies for recovery in PFASST is straightforward but does not allow for meaningful runtime measurements.
However, for tests on actual HPC machines using a lower-level programming language like C++ and advanced communication libraries like MPI, a number of technical questions arise.
In particular, analysing the impact of start-up times for replacement processes will be critical, since it could constitute an important bottleneck.
Also, realistic hard-faults are more complex than the simulated faults used in the paper.

For the transition to non-emulated faults, a suitable MPI implementation is necessary that robustly detects and deals with non-responsive MPI processes during sends and receives.
It also has to provide an interface for user-level recovery procedures and a reliable mechanism to replace or restart a failed process.
This replacement process could be the one that failed after a reboot, a spare process on a different node or one of the other time-parallel processes.
In the latter case, information from one of the other time steps has to be dropped in order to take over the failed one.
If one of the earlier time steps has already converged, the choice is easy. 
If all processes are still iterating, on the other hand, using the very last process as replacement seems to be reasonable, since here the iteration is in the earliest stage.
In contrast to the fault emulation presented here, a parallel implementation furthermore necessitates a number of additional technical steps during recovery: data from adjacent processes have to be received and these processes have to know that they are expected to send these data.
Then, the problem has to be set up on the replacement process, including allocation of memory.
Finally, for the identification as well as for the restart of a replacement process, some kind of supervisor or at least global, up-to-date information on all processes seems to be necessary.
Efficient implementation of recovery strategies into the MPI-based C++-framework PFASST++~\cite{Klatt2015} is ongoing work.

\paragraph{Convergence analysis}
In most of the cases shown in this work, two-sided interpolation with coarse correction seemed to be among the most effective strategies, but in others the coarse correction provided no benefit at all.
As mentioned in the numerical experiments of Section~\ref{sec:experiments}, the quality of the coarse level is essential for the efficiency of the different recovery strategies.
To predict the effect of faults, different recovery strategies and additional coarse corrections as well as to identify optimal strategies for a given setup, an in-depth mathematical analysis of PFASST is required.
Unfortunately, such an analysis proves to be difficult and is the subject of substantial ongoing efforts.
A recent paper casts the method into the framework of space-time multigrid methods \cite{BoltenEtAl2016} and hopefully this will allow for a comprehensive analysis in the future.

\section{Summary}
We present and compare different strategies that could, in principle, allow the parallel-in-time integration method PFASST to recover from hard faults and subsequent loss of data.
As a \emph{parallel-across-the-steps} method, PFASST stores solutions at multiple points in time on different processors, which allows to recover a solution that is lost due to process failure.
A theoretical model links the overhead of different approaches to the number of additional iterations required by fault-tolerant PFASST to reach a specific residual tolerance.
Efficiency of the different strategy is assessed in multiple examples of diffusive and advective type.
For both the Gray-Scott reaction diffusion model and the Boussinesq model for compressible flows we demonstrate that fault-tolerant PFASST can allow a simulation to make progress even when subjected to a high number of randomly occurring faults.
Since PFASST shares features with other parallel-in-time methods like Parareal or MGRIT, similar strategies to exploit algorithm-based fault tolerance for parallel-in-time integration could be devised for other methods as well.
The paper focusses on assessing the impact of simulated faults and subsequent recovery on the convergence of PFASST and leaves runtime measurements in a parallel environment for future work.

\section*{Acknowledgments}
The outline of this paper was drafted during a meeting of both authors at the Centre for Interdisciplinary Research (ZIF) at the University of Bielefeld in Germany.
D. Ruprecht thankfully acknowledges an invitation by the ZIF for a research visit in August 2015 during which this meeting took place.

\end{document}